%% file: main.tex
\documentclass[10pt, conference, letterpaper]{IEEEtran}
\IEEEoverridecommandlockouts
\usepackage{cite}
\usepackage{amsfonts,amsmath,color,amssymb,amsxtra,amsbsy,floatflt}
\usepackage{algorithmic}
\usepackage{graphicx}
\usepackage{textcomp}
\usepackage{xcolor}
\def\BibTeX{{\rm B\kern-.05em{\sc i\kern-.025em b}\kern-.08em
    T\kern-.1667em\lower.7ex\hbox{E}\kern-.125emX}}

\usepackage{enumitem}
\newlist{myenum}{enumerate}{3}
\setlist[myenum,1]{label*=\arabic*)}
\setlist[myenum,2]{label=\arabic{myenumi}.\arabic*)}
\setlist[myenum,3]{label=\arabic{myenumi}.\arabic{myenumii}.\arabic*)}

\def\1{\mathbf{1}}

\usepackage[acronym]{glossaries}
\usepackage{subfigure}
\usepackage[ruled,vlined,linesnumbered]{algorithm2e}
\usepackage{placeins}
\usepackage{epstopdf}
\usepackage{enumitem}

 \usepackage{tikz}
\usetikzlibrary{automata,arrows,positioning,calc}

\usepackage{multirow}
\usepackage[export]{adjustbox}
\usepackage{xfrac}
\usepackage[hidelinks]{hyperref}

\newcommand{\smallminus}{\scalebox{0.5}[1.0]{$-$}}

\newcommand{\name}{ORANUS\xspace}
\renewcommand{\baselinestretch}{0.97}

\usepackage{fancyhdr}

\begin{document}


\pagestyle{fancy}
\fancyhf{} 
\renewcommand{\headrulewidth}{0pt}
\fancyhead{} 
\fancyhead[C]{\fontsize{8}{10} \selectfont This article has been accepted for publication in the IEEE International Conference on Computer Communications (INFOCOM), 2024. } 

\input{acronyms.tex}

\title{\name : Latency-tailored Orchestration via Stochastic Network Calculus in 6G O-RAN\\
\thanks{The research leading to these results has been supported in part by SNS JU Project 6G-GOALS (GA no. $101139232$), in part by the EU FP Horizon 2020 project DAEMON (GA no. $101017109$), in part by the Spanish Ministry of Economic Affairs and Digital Transformation under Projects 6G-CHRONOS and OPEN6G (grants TSI-063000-2021-28 and TSI-063000-2021-3) and in part by the Spanish Ministry of Science and Innovation / State Investigation Agency under Project 6G-INSPIRE (grant PID$2022$-$137329$OB-C$43$).}
}


\author{\IEEEauthorblockN{
Oscar Adamuz-Hinojosa\IEEEauthorrefmark{1},
Lanfranco Zanzi\IEEEauthorrefmark{2},
Vincenzo Sciancalepore\IEEEauthorrefmark{2}, \\
Andres Garcia-Saavedra\IEEEauthorrefmark{2}, 
Xavier~Costa-Pérez\IEEEauthorrefmark{3}\IEEEauthorrefmark{2}\IEEEauthorrefmark{4}
}

\IEEEauthorblockA{
    \IEEEauthorrefmark{1}Department of Signal Theory, Telematics and Communications, University of Granada, Spain. Email: oadamuz@ugr.es,\\ }
\IEEEauthorblockA{\IEEEauthorrefmark{2} NEC Laboratories Europe, Heidelberg, Germany.  Email:\{name.surname\}@neclab.eu,\\ }
 \IEEEauthorrefmark{3}i2CAT Foundation, Barcelona, Spain.  Email:\{name.surname\}@i2cat.net,
\IEEEauthorblockA{  \IEEEauthorrefmark{4} ICREA, Barcelona, Spain.}
}

\maketitle

\begin{abstract}
The Open Radio Access Network (O-RAN)-compliant solutions lack crucial details to perform effective control loops at multiple time scales. In this vein, we propose \name, an O-RAN-compliant mathematical framework to allocate radio resources to multiple ultra Reliable Low Latency Communication (uRLLC) services. 
In the near-RT control loop, \name relies on a novel \emph{Stochastic Network Calculus (SNC)-based model} to compute the amount of guaranteed radio resources for each uRLLC service. Unlike traditional approaches as queueing theory, the SNC-based model allows \name to ensure the probability the packet transmission delay exceeds a budget, i.e., the violation probability, is below a target tolerance. \name also utilizes a RT control loop to monitor service transmission queues, dynamically adjusting the guaranteed radio resources based on detected traffic anomalies. To the best of our knowledge, \name is the first O-RAN-compliant solution which benefits from SNC to carry out near-RT and RT control loops. Simulation results show that \name significantly improves over reference solutions, with an average violation probability $10\times$ lower.
\end{abstract}


\section{Introduction}
In the 
\gls{6G} networks, a pivotal scenario to address is the coexistence of multiple \gls{uRLLC} services.
They place stringent demands on latency and reliability, requiring deterministic
guarantees to ensure their seamless operation~\cite{Popovski2018}. Moreover, a key driving factor in \gls{6G} networks is the virtualization of the \gls{RAN}~\cite{Tang2023}. This entails the deployment of \glspl{vRAN} instances, wherein each \gls{vRAN} comprises a set of fully-configurable \glspl{vBS} designed to cater the 
requirements of individual services. 
In this context, the \gls{O-RAN} Alliance 
proposed a novel architecture~\cite{Abdalla2022} embracing and promoting the \gls{3GPP} functional split, where each \gls{vBS} is divided across multiple network nodes: \gls{CU}-\gls{CP}, \gls{CU}-\gls{UP}, \gls{DU} and \gls{RU}. Furthermore, the \gls{O-RAN} architecture considers two \glspl{RIC}, which provide a centralized abstraction of the network, allowing the \gls{MNO} to perform autonomous actions between \gls{vBS} components and their controllers. Specifically, the non-\gls{RT} \gls{RIC} supports large timescale optimization tasks (i.e., in the order of seconds or minutes), including policy computation and \gls{ML} model management. Such functionalities are carried out by third-party applications denominated \emph{rApps}. Additionally, the near-\gls{RT} \gls{RIC} performs \gls{RAN} optimization, control and data monitoring tasks in near-\gls{RT} timescales (i.e., from $10$ms to $1$s). Such functionalities can also be performed by third-party applications denominated \emph{xApps}. For more details about \gls{O-RAN}, we recommend~\cite{survey-O-RAN}.




Despite the ongoing standardization efforts, there are still open challenges toward a successful implementation of the \gls{O-RAN} architecture: limiting the execution of control tasks in both \glspl{RIC} prevents the use of solutions where decisions must be made in \emph{real-time}, i.e., below 10 ms~\cite{dApps_article}. For example, \gls{uRLLC} \gls{MAC} scheduling requires making decisions at sub-millisecond timescales~\cite{LACO}.
Unfortunately, the near-\gls{RT}-\gls{RIC} might struggle to accomplish this procedure due to limited access to low-level information (e.g., transmission queues, channel quality, etc.). The potential high latency involved in obtaining this information further exacerbates the problem. 
This calls for a \gls{RT} control loop to monitor and orchestrate 
the decision of 
\gls{MAC} schedulers.
Addressing the current mention of \gls{RT} control loop standardization as a study item ~\cite{dApps_article}, in this paper we provide the basis for future research activities towards an \gls{RT} orchestration framework.

In this context, an important yet unaddressed challenge lies in coordinating different control loops, which operate at different time scales 
~\cite{survey-O-RAN}. The need for seamless and reliable coexistence of diverse \gls{uRLLC} services demands, effective coordination schemes between the near-RT and non-RT control loops, as well as requires mechanisms to align the decisions made by different control loops while optimizing the resource allocation and avoiding conflicts.
In this paper, we advocate for the adoption of \gls{SNC} to model the complex dynamics of O-RAN systems and analyze the performance of communications in terms of violation probability, i.e., the probability of a packet being transmitted exceeding a delay bound, while considering the uncertainties and variability in traffic patterns and channel conditions.
Several works~\cite{SOTANC1,SOTANC2,SOTANC3} already leveraged \gls{SNC} 
to estimate delay bounds in the radio interface for given target tolerance and specific resource allocation in single \gls{uRLLC} services. 
Conversely, in~\cite{Adamuz-Hinojosa-TWC2023} the authors extend the application of \gls{SNC} to resource planning of multiple \gls{uRLLC} services. However, their approach assumes dedicated RBs per service, leading to 
potential resource wastage.
Regarding the \gls{RB} allocation at \gls{RT} scale, the existing literature is vast. However, to the best of our knowledge, there are no \gls{RT} solutions that benefit from \gls{SNC} models.

{\bf Contributions.} In this work, we focus our mathematical discussion and empirical evaluation on the \gls{DL} operation of a single cell supporting multiple \gls{uRLLC} services, each one with specific requirements in terms of packet delay budget and violation probability. However, our solution can be readily applied to extended scenarios, such as \gls{UL} transmissions and multiple cells. The main contributions are:
\begin{itemize}
    \item[(C1)] We propose \name, an \gls{O-RAN}-compliant mathematical framework to carry out the multi-time-scale control loops for the radio resource allocation to multiple \gls{uRLLC} services, focusing on near-\gls{RT} and \gls{RT} scales.
    \item[(C2)] To perform the near-\gls{RT} control loop, \name relies on a novel \gls{SNC}-based controller to compute the amount of guaranteed \glspl{RB} for each \gls{uRLLC} service, which ensures them the violation probability is below a target tolerance. Additionally, the proposed \gls{SNC}-based controller can directly use real metrics of the incoming traffic and channel conditions to capture their statistical distributions. 
    \item[(C3)] Considering the amount of guaranteed \glspl{RB} computed at near-\gls{RT} scale, we propose a \gls{RT} control loop to monitor the transmission queue of each service. If traffic anomalies are detected, the proposed control loop adapts accordingly the amount of guaranteed \glspl{RB} for the corresponding services, as to mitigate the violation probability.
\end{itemize}

To the best of our knowledge, \name is the first \gls{O-RAN}-based solution which uses \gls{SNC} to perform the \gls{RB} allocation to multiple \gls{uRLLC} services at near-\gls{RT} scale while a \gls{RT} control loop adapts such allocation in response to traffic anomalies.

The remainder of this paper is organized as follows. Section~\ref{sec:OurFramework} defines \name. Then, we present the proposed \gls{SNC} model in Section~\ref{sec:SNCmodel}. In Section~\ref{sec:SNC-basedOrchestrator}, we explain how  \name performs the multi-time-scale control loops. Section~\ref{sec:Results} evaluates the performance of \name. Section~\ref{sec:RelatedWorks} discusses the related works. Finally, Section~\ref{sec:Conclusions} concludes this paper.


\section{The \name Framework}\label{sec:OurFramework}
\begin{figure}[t!]
    \centering
    \includegraphics[width=0.75\columnwidth]{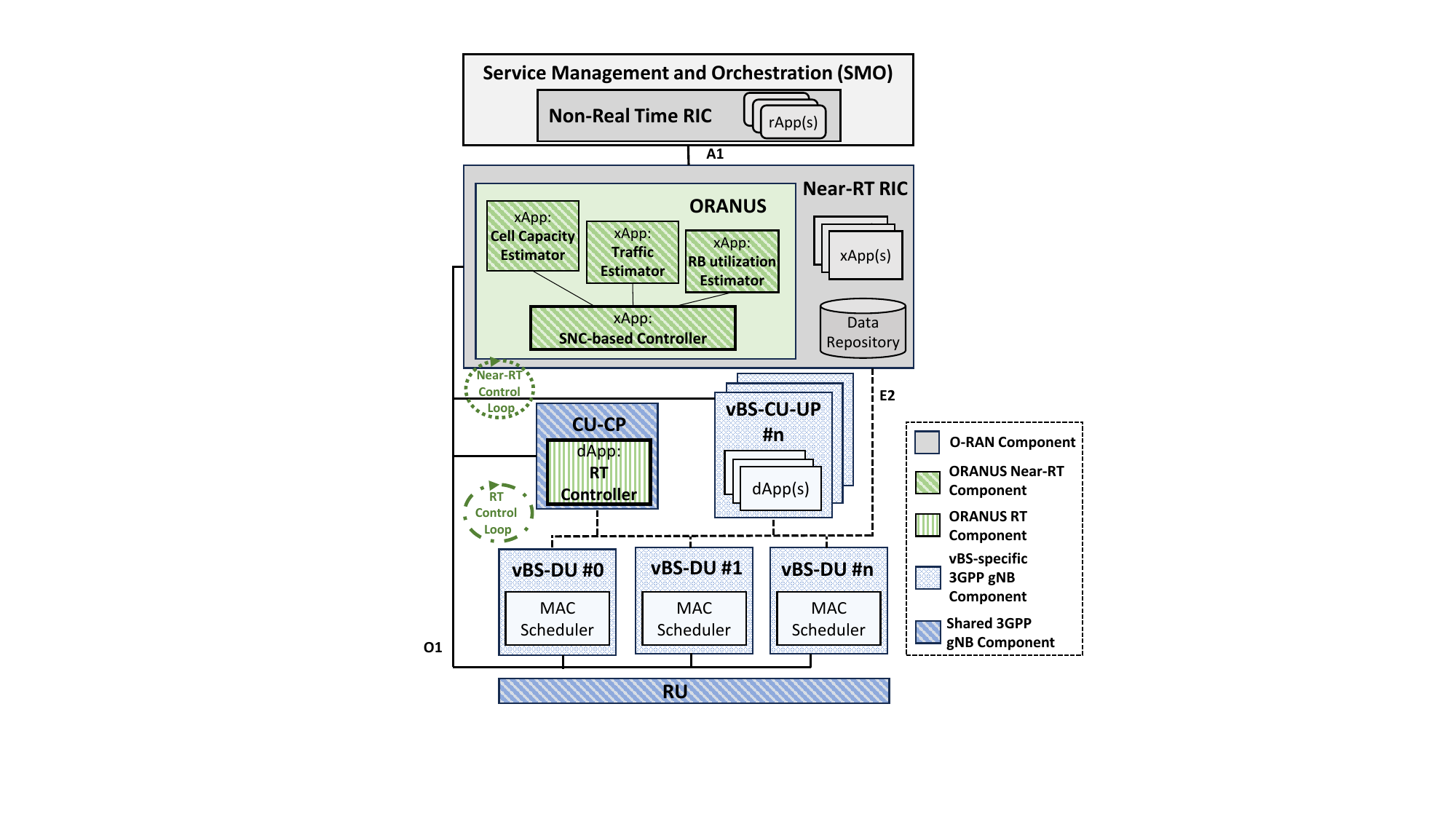}
    \caption{Integration of \name (i.e., green blocks) in the O-RAN architecture.}
    \label{fig:FrameworkIntegration}
\end{figure}

Following the O-\gls{RAN} specifications~\cite{O-RAN-workflow,O-RAN-workflow2}, Fig.~\ref{fig:FrameworkIntegration} depicts the main functional blocks of \name. Specifically, \name comprises five \emph{xApps} and one \emph{dApp}, as initially proposed by~\cite{dApps_article}. The Cell Capacity Estimator, Traffic Estimator, \gls{RB} utilization Estimator and \gls{SNC}-based Controller \emph{xApps} are located in the near-\gls{RT} \gls{RIC} and they are responsible for the \gls{RB} allocation of multiple \gls{uRLLC} services in a near-\gls{RT} scale. The \gls{RT} Controller \emph{dApp} is located in a \gls{CU}-\gls{CP}, which is shared by the \gls{uRLLC} services deployed in the same cell\footnote{We interchangeably use the terms \gls{uRLLC} service and \gls{vBS} because we assume each \gls{uRLLC} service is deployed using a specific \gls{vBS} within a cell.} and it is responsible for controlling the \gls{RB} allocation at \gls{RT} scale for these services. We also assume the \gls{CU}-\gls{UP} and the \glspl{DU} are dedicated per service. Below is a brief summary of the tasks performed by these \emph{Apps}. 

\textbf{SNC-based Controller \emph{xApp}}: It obtains the guaranteed amount of \glspl{RB} $N_{\scriptscriptstyle m}^{\scriptscriptstyle min}$ for each service $m$ $\in \mathcal{M}$, which ensures $\text{P}\left[w>W_{\scriptscriptstyle m}^{\scriptscriptstyle th}\right]<\varepsilon_{\scriptscriptstyle m}$, where $w$ is the packet transmission delay\footnote{\emph{Packet transmission delay} is the waiting time of a Transport Block (TB) unit from entering the transmission buffer until it is fully transmitted.}, $W_{\scriptscriptstyle m}^{\scriptscriptstyle th}$ is the delay budget and $\varepsilon_{\scriptscriptstyle m}$ is the target violation probability. To this end, it operates every $T_{ \scriptscriptstyle OUT}$ \glspl{TTI} and exchanges information with the Traffic/Cell Capacity Estimator \emph{xApps} and the \gls{RB} utilization Estimator \emph{xApp}.

\textbf{Traffic and Cell Capacity Estimator \emph{xApps}}: 
They analyze each service $m$ $\in \mathcal{M}$ and estimate the incoming traffic demand in terms of the number of bits per \gls{TTI}.
Additionally, they compute arrays of samples, each representing the number of bits that each cell can accommodate for these services in an arbitrary \gls{TTI}, based on the chosen \glspl{MCS} for packet transmission.
To that end, these \emph{xApps} collect these metrics from each DU via the O1 interface.

\textbf{\gls{RB} utilization Estimator \emph{xApp}}: It estimates the \gls{PMF} of the \gls{RB} utilization. To that end, it relies on a \gls{NN} based on \gls{MDN}. It considers the following inputs: (a) the incoming traffic demand in each \gls{TTI} expressed in bits, (b) the enqueued bits in each \gls{TTI} and (c) the candidate number of guaranteed \glspl{RB} for the next $T_{\scriptscriptstyle OUT}$ \glspl{TTI}. These metrics are available from E2/O1 interfaces. 

\textbf{\gls{RT} Controller \emph{dApp}}: It operates every \gls{TTI} and is responsible for ensuring each DU MAC scheduler, one per service $m$ $\in \mathcal{M}$, has available at least $N_{\scriptscriptstyle m}^{\scriptscriptstyle min}$ \glspl{RB}. Note that $\sum_{\scriptscriptstyle m \in \mathcal{M}}N_{\scriptscriptstyle m}^{\scriptscriptstyle min}  \leq N_{\scriptscriptstyle cell}^{\scriptscriptstyle RB}$, where $N_{\scriptscriptstyle cell}^{\scriptscriptstyle RB}$ is the \glspl{RB} available in the cell. Focusing on a single \gls{TTI} $t$, if a service $m$ requires a number of \glspl{RB} $N_{\scriptscriptstyle t}$ such as $N_{\scriptscriptstyle t}<N_{\scriptscriptstyle m}^{\scriptscriptstyle min}$, this \emph{dApp} will allocate $N_{\scriptscriptstyle t}$ \glspl{RB} to such service. 
Otherwise, it will first check if there are any available free \glspl{RB}, i.e., those \glspl{RB} that were initially allocated to other services but have not been used in the current \gls{TTI}. If free \glspl{RB} are found, the \emph{dApp} allocates $N_{\scriptscriptstyle m}^{\scriptscriptstyle min}$ plus the available free \glspl{RB} to the specified service. Additionally, this \emph{dApp} monitors the transmission buffers to detect traffic anomalies. If so, it temporarily updates $N_{\scriptscriptstyle m}^{\scriptscriptstyle min}$ $\forall m \in \mathcal{M}$ to mitigate the violation probability. 

\section{Latency Model based on SNC}\label{sec:SNCmodel}
To assess the packet transmission delay, we propose a \gls{SNC} model. In this section, we first provide some fundamentals on \gls{SNC}. Then, we describe the steps to compute the delay bound $W$. Finally, we particularize them to our scenario.

\subsection{Fundamentals on SNC}
We use \gls{SNC} to model the incoming and outgoing traffic of a network node by stochastic arrival and service processes. The arrival process $A(\tau,t)$ represents the cumulative number of bits that arrive at this node, and the service process $S(\tau,t)$ denotes the cumulative number of bits that may be served by this node. Both measured over the time interval $(\tau,t]$.  

\gls{SNC} relies on the \gls{EBB} and the \gls{EBF} models~\cite{Fidler1}. They define an upper bound $\alpha (\tau,t)$, denominated arrival envelope, and a lower bound $\beta  (\tau,t)$, denominated service envelope, as Eqs.  \eqref{eq:EBBAffine} and \eqref{eq:EBF_Service_Envelope} show. The parameters $\varepsilon_A$ and $\varepsilon_S$ are known as the overflow and deficit profiles.
\begin{equation}
   \text{P}[A(\tau,t) > \alpha(\tau,t)] \leq \varepsilon_{\scriptscriptstyle A}.
    \label{eq:EBBAffine}
\end{equation}
\begin{equation}
    \text{P}\left[ S(\tau,t) < \beta(\tau,t) \right] \leq \varepsilon_{\scriptscriptstyle S}.
    \label{eq:EBF_Service_Envelope}
\end{equation}

A widely accepted practice in \gls{SNC} consists of assuming affine functions to define $\alpha (\tau,t)$ and $\beta  (\tau,t)$. These functions are defined in Eqs.~\eqref{eq:AffineArrivalEnvelope} and \eqref{eq:ServiceEnvelopeBeta}, where the parameters $\rho_{\scriptscriptstyle A} > 0$, $\rho_{\scriptscriptstyle S} > 0$ and $b_{\scriptscriptstyle A} \geq 0$, $b_{\scriptscriptstyle S} \geq 0$ are the rate and burst parameters for $\alpha (\tau,t)$ and $\beta (\tau,t)$, respectively. Additionally, $\left[x \right]_{+}$ denotes $\text{max}\{0,x \}$. Finally, $\delta> 0$ is a sample path argument considered by the \gls{EBB} and the \gls{EBF} models~\cite{Fidler1}.
\begin{equation}
    \alpha (\tau,t) = (\rho_{\scriptscriptstyle A}+\delta)[t-\tau] + b_{\scriptscriptstyle A}.
    \label{eq:AffineArrivalEnvelope}
\end{equation}
\begin{equation}
    \beta (\tau,t)= (\rho_{\scriptscriptstyle S} -\delta)\left[t-\tau -\sfrac{b_{\scriptscriptstyle S}}{\rho_{\scriptscriptstyle S}} \right]_{+}.
    \label{eq:ServiceEnvelopeBeta}
\end{equation}

If we represent $\alpha(\tau,t)$ and $\beta(\tau,t)$ over a $t-\tau$ axis, we can compute the delay bound $W$ as the horizontal deviation between these envelopes. Specifically, this horizontal deviation can be formulated as Eq. \eqref{eq:Delay_bound_v2} shows. Note that this deviation exists when the slope of $\beta (\tau,t)$ is greater than the slope of $\alpha(\tau,t)$. This results in the condition defined in Eq. \eqref{eq:ConditionDelay}.
\begin{equation}
    W = \tfrac{b_{\scriptscriptstyle A} + b_{\scriptscriptstyle S}}{\rho_{\scriptscriptstyle S}-\delta}.
    \label{eq:Delay_bound_v2}
\end{equation}
\begin{equation}
    \rho_{\scriptscriptstyle S} - \delta  > \rho_{\scriptscriptstyle A} + \delta.
    \label{eq:ConditionDelay}
\end{equation}


\subsection{SNC Steps to Compute the Delay Bound}
Below, we briefly summarize the steps proposed in~\cite[Sections II.A and II.B]{Fidler1} to obtain the arrival envelope, the service envelope and the delay bound.
\begin{enumerate}
    \item We compute the \glspl{MGF} for the arrival and service processes, i.e., $M_{\scriptscriptstyle A}(\theta)$ and $M_{\scriptscriptstyle S}(\smallminus\theta)$. The \gls{MGF} of a random process $X$ is defined as $\text{E}\left[e^{\theta X}\right]$ with free parameter $\theta$. Note that searching a lower bound for the service process is equivalent to using the sign $(\smallminus)$ for $\theta$.
    \item We define upper bounds for the \glspl{MGF} as Eqs. \eqref{eq:ArrivalAffineEnvelopeModel} and \eqref{eq:AffineEnvelopeServiceMGF} show. They are characterized by the rate parameters $\rho_{\scriptscriptstyle A}$ and $\rho_{\scriptscriptstyle S}$, and the burst parameters $\sigma_{\scriptscriptstyle A}$ and $\sigma_{\scriptscriptstyle S}$. Note that $\rho_{\scriptscriptstyle A}$ and $\rho_{\scriptscriptstyle S}$ are the same as the ones defined in Eqs. \eqref{eq:AffineArrivalEnvelope} and \eqref{eq:ServiceEnvelopeBeta}. If we replace the left side of Eqs. \eqref{eq:ArrivalAffineEnvelopeModel} and \eqref{eq:AffineEnvelopeServiceMGF}  by the \glspl{MGF} computed in the previous step, we can obtain the parameters $\rho_{\scriptscriptstyle A}$,  $\sigma_{\scriptscriptstyle A}$, $\rho_{\scriptscriptstyle S}$ and $\sigma_{\scriptscriptstyle S}$.
    \begin{equation}
        M_{\scriptscriptstyle A}(\theta) \leq \text{exp}\left[ \theta \left(\rho_{\scriptscriptstyle A}[t-\tau] + \sigma_{\scriptscriptstyle A} \right) \right].
        \label{eq:ArrivalAffineEnvelopeModel}
    \end{equation}
    \begin{equation}
        M_{\scriptscriptstyle S}(\smallminus\theta) \leq \text{exp}\left[ \smallminus\theta \left(\rho_{\scriptscriptstyle S}[t-\tau] -\sigma_{\scriptscriptstyle S} \right) \right].
        \label{eq:AffineEnvelopeServiceMGF}
    \end{equation}
    \item The \gls{EBB} and \gls{EBF} models defined in Eqs. \eqref{eq:EBBAffine} and \eqref{eq:EBF_Service_Envelope}, and the \glspl{MGF} are directly connected by the Chernoff bound~\cite{ross2014first}. Based on it, we can obtain $b_{\scriptscriptstyle A} $ and $b_{\scriptscriptstyle S} $ as Eqs. \eqref{eq:Burst_parameter_arrival} and \eqref{eq:burst_parameter_service} shows. Note that a common practice is to equally distribute the target violation probability among the overflow and deficit profiles, i.e., $\varepsilon_{\scriptscriptstyle A}=\varepsilon_{\scriptscriptstyle B}=\varepsilon/2$.
    \begin{equation}
        b_{\scriptscriptstyle A} = \sigma_{\scriptscriptstyle A} - \tfrac{1}{\theta} \left[\text{ln}(\varepsilon_{\scriptscriptstyle A}) + \text{ln}\left( 1-\text{exp}\left[ \smallminus\theta \delta\right] \right) \right].
        \label{eq:Burst_parameter_arrival}
    \end{equation}
    \begin{equation}
        b_{\scriptscriptstyle S} = \sigma_{\scriptscriptstyle S} - \tfrac{1}{\theta} \left[\text{ln}(\varepsilon_{\scriptscriptstyle S}) + \text{ln}\left( 1-\text{exp}\left[\smallminus\theta \delta \right]\right) \right].
        \label{eq:burst_parameter_service}
    \end{equation}
\item Using the previous results in Eq. \eqref{eq:Delay_bound_v2}, we can rewrite the delay bound $W$ as Eq. \eqref{eq:Middle_delay_bound} shows. Then, we need to search the values for $\theta$ and $\delta$ that minimize the expression of Eq. \eqref{eq:Middle_delay_bound} to estimate the delay bound.
\begin{equation}
    W = \frac{ \sigma_{\scriptscriptstyle A}+\sigma_{\scriptscriptstyle S}-\tfrac{2}{\theta}\left[ \text{ln} \left(\frac{\varepsilon}{2} \right) + \text{ln} \left(1-\text{exp}\left[-\theta\delta\right] \right)\right] }{\rho_{\scriptscriptstyle S} -\delta}.
    \label{eq:Middle_delay_bound}
\end{equation}
\end{enumerate}

Below, we particularize the arrival process $A(\tau,t)$, the service process $S(\tau,t)$ and the steps to estimate $W$  for a scenario where a cell implements an \gls{uRLLC} service.

\subsection{uRLLC Traffic Model}\label{sec:uRLLCTrafficModel}
The arrival process $A_{\scriptscriptstyle m}(\tau,t)$ is the cumulative \gls{DL} traffic of the service $m$, which is defined in Eq. \eqref{eq:ArrivalProcesSlice}. The variable $N_{\scriptscriptstyle slot}(\tau,t) = (t-\tau)/t_{\scriptscriptstyle slot}$ represents the total number of \glspl{TTI} in $(\tau,t]$, and $t_{\scriptscriptstyle slot}$ denotes the duration of a \gls{TTI}. The random variable $D_{\scriptscriptstyle m}(i)$ represents the number of bits that arrive to the transmission buffer for service $m$ in the $i$-th \gls{TTI}.
\begin{equation}
    A_{\scriptscriptstyle m}(\tau,t)=\sum_{i=\tau}^{N_{\scriptscriptstyle slot}(\tau,t)} D_{\scriptscriptstyle m}(i).
    \label{eq:ArrivalProcesSlice}
\end{equation}

We assume that the \gls{PMF} of $D_{\scriptscriptstyle m}(i)$ can be estimated by using samples of the incoming bits per \gls{TTI} in the last $T_{\scriptscriptstyle OBS}$ \glspl{TTI}. Specifically, we define the sample vector $\vec{x}_{\scriptscriptstyle D_{\scriptscriptstyle m}} =\{d_{\scriptscriptstyle m,1}^{\scriptscriptstyle in}, d_{\scriptscriptstyle m,2}^{\scriptscriptstyle in}\, \hdots\, d_{\scriptscriptstyle m,T_{\scriptscriptstyle OBS}}^{\scriptscriptstyle in} \}$, where $d_{\scriptscriptstyle m,i}^{\scriptscriptstyle in}$ denotes the number of bits that arrive to the transmission buffer in the \gls{TTI} $i$ for the service $m$. Additionally, $d_{\scriptscriptstyle m,i}^{\scriptscriptstyle in}=\sum_{j=1}^{J_{\scriptscriptstyle m,i}^{\scriptscriptstyle in}}l_j$, where $J_{\scriptscriptstyle m,i}^{\scriptscriptstyle in}$ is the number of incoming packets for service $m$ in the \gls{TTI} $i$ and $l_{\scriptscriptstyle j}$ the size of the packet $j$. Note that the computation of $\vec{x}_{\scriptscriptstyle D_{\scriptscriptstyle m}}$ is a task performed by the Traffic Estimator \emph{xApp}.
With these assumptions, we can compute the \gls{MGF} for $D_{\scriptscriptstyle m}(i)$: 
\begin{equation}
    M_{\scriptscriptstyle D_{\scriptscriptstyle m}}(\theta) =\frac{1}{T_{\scriptscriptstyle OBS}} \sum_{i=1}^{T_{\scriptscriptstyle OBS}} \text{exp}\left[\theta d_{\scriptscriptstyle m,i}^{\scriptscriptstyle in}\right].
    \label{eq:MGFTraffic}
\end{equation}

Considering $M_{\scriptscriptstyle D_{\scriptscriptstyle m}}(\theta)$, we can compute the \gls{MGF} of the arrival process $A_{\scriptscriptstyle m}(\tau,t)$ as Eq. \eqref{eq:MGFTraffic2} shows.
\begin{equation}
            M_{\scriptscriptstyle A_{\scriptscriptstyle m}}(\theta)   = e^{\left[\text{ln}\left( \tfrac{1}{T_{\scriptscriptstyle OBS}} \sum_{i=1}^{T_{\scriptscriptstyle OBS}} \text{exp}\left[\theta d_{\scriptscriptstyle m,i}^{\scriptscriptstyle in}\right]\right) \tfrac{ (t-\tau)}{t_{\scriptscriptstyle slot}}\right]}.
    \label{eq:MGFTraffic2}
\end{equation}

Finally, by equaling the left and right sides of Eq. \eqref{eq:ArrivalAffineEnvelopeModel}, we obtain $\rho_{\scriptscriptstyle A_{\scriptscriptstyle m}}$ as Eq. \ref{Eq:ArrivalEnvelopeParameters} shows. Note that $\sigma_{\scriptscriptstyle A_{\scriptscriptstyle m}}=0$.
\begin{equation}
        \rho_{\scriptscriptstyle A_{\scriptscriptstyle m}} = \frac{\text{ln}\left[\tfrac{1}{T_{\scriptscriptstyle OBS}} \sum_{i=1}^{T_{\scriptscriptstyle OBS}} \text{exp}\left[\theta d_{\scriptscriptstyle m,i}^{\scriptscriptstyle in}\right]\right] }{\theta t_{\scriptscriptstyle slot}}.  
\label{Eq:ArrivalEnvelopeParameters}
\end{equation}

\subsection{Service Model}\label{sec:ServiceModel}
The service process $S_{\scriptscriptstyle m}(\tau,t)$ represents the accumulated capacity the cell may provide to the service $m$ and can be defined as
\begin{equation}
S_{\scriptscriptstyle m}(\tau,t) = \sum_{i=\tau}^{N_{\scriptscriptstyle slot}(\tau, t)} C_{\scriptscriptstyle m}(i),
    \label{eq:ServiceCurveRANSlice}
\end{equation}
where the random variable $C_{\scriptscriptstyle m}(i)$ is the number of bits that may be served by the cell for service $m$ in the \gls{TTI} $i$.

The negative \gls{MGF} for $C_{\scriptscriptstyle m}(i)$ is defined in Eq. \eqref{eq:MGFAvailableCapacityforSlice}. To that end, we consider the \gls{PMF} of $C_{\scriptscriptstyle m}(i)$ can be estimated by (a) using samples of the number of bits which may be transmitted in the last $T_{\scriptscriptstyle OBS}$ \glspl{TTI} and (b) considering the \gls{PMF} of the \gls{RB} utilization. The latter captures the dynamics of the \gls{RT} Controller \emph{dApp}. Specifically, we consider the sample vectors $\vec{x}_{C_m,n}  =\{c_{\scriptscriptstyle m,1}^{\scriptscriptstyle n,out}, c_{\scriptscriptstyle m,2}^{\scriptscriptstyle n,out}\, \hdots\, c_{\scriptscriptstyle m,T_{\scriptscriptstyle m,n}}^{\scriptscriptstyle n,out} \}$  $\forall n \in [0,\; 
N_{\scriptscriptstyle add}]$ where $c_{\scriptscriptstyle m,i}^{\scriptscriptstyle n,out}$ denotes the number of bits which may be transmitted by the cell for service $m$ in a single \gls{TTI}, i.e., considering the serving cell uses $n + N_{\scriptscriptstyle m}^{\scriptscriptstyle min}$ \glspl{RB} for such service. Additionally, $N_{\scriptscriptstyle add} = N_{\scriptscriptstyle cell}^{\scriptscriptstyle RB}-N_{\scriptscriptstyle m}^{\scriptscriptstyle min}$, where $T_{\scriptscriptstyle m,n}$ is the number of samples. Additionally, we consider $\pi_{\scriptscriptstyle m,n}$ as the probability the service $m$ has available  $n+N_{\scriptscriptstyle m}^{\scriptscriptstyle min}$ \glspl{RB} in an arbitrary \gls{TTI}, conditioned to this service $m$ needs more \glspl{RB} than $N_{\scriptscriptstyle m}^{\scriptscriptstyle min}$. The computation of $\vec{x}_{\scriptscriptstyle C_{\scriptscriptstyle m},n}$, a task performed by the Cell Capacity Estimator \emph{xApp}, is detailed in Section \ref{sec:CellCapacityEstimator}. 
\begin{equation}
\begin{split}
    M_{\scriptscriptstyle C_{\scriptscriptstyle m}}(\smallminus \theta) =\sum_{n=0}^{
N_{\scriptscriptstyle add}} \tfrac{\pi_{\scriptscriptstyle m,n}}{T_{\scriptscriptstyle m,n}} \sum_{i=1}^{T_{\scriptscriptstyle m,n}} \text{exp} \left[ \smallminus \theta c_{\scriptscriptstyle m,i}^{\scriptscriptstyle n,out}\right].
\end{split}
    \label{eq:MGFAvailableCapacityforSlice}
\end{equation}

Considering $M_{\scriptscriptstyle C_{\scriptscriptstyle m}}(\smallminus\theta)$, we can compute the negative \gls{MGF} of the service process $S_{\scriptscriptstyle m}(\tau,t)$ as
\begin{equation}
            M_{\scriptscriptstyle S_{\scriptscriptstyle m}}(\smallminus\theta)  = e^{\left[ \text{ln}\left(  \sum_{n=0}^{
N_{\scriptscriptstyle add}} \tfrac{\pi_{\scriptscriptstyle m,n}}{T_{\scriptscriptstyle m,n}} \sum_{i=1}^{T_{\scriptscriptstyle m,n}} \text{exp} \left[ \smallminus \theta c_{\scriptscriptstyle m,i}^{\scriptscriptstyle n,out}\right]\right) \frac{(t-\tau)}{t_{slot}} \right] }.
    \label{eq:MGFranslice1}
\end{equation}
Finally, by equaling the left and right sides of Eq. \eqref{eq:AffineEnvelopeServiceMGF}, we obtain $\rho_{\scriptscriptstyle S_{\scriptscriptstyle m}}$ as Eq. \ref{Eq:ServiceEnvelopeParameters} shows. Note that $\sigma_{\scriptscriptstyle S_{\scriptscriptstyle m}}=0$.
\begin{equation}
        \rho_{\scriptscriptstyle S_{\scriptscriptstyle m}} =  \frac{-\text{ln}\left[  \sum_{n=0}^{
N_{\scriptscriptstyle add}} \frac{\pi_{\scriptscriptstyle m,n}}{T_{\scriptscriptstyle m,n}} \sum_{i=1}^{T_{\scriptscriptstyle m,n}} \text{exp} \left[ \smallminus \theta c_{\scriptscriptstyle m,i}^{\scriptscriptstyle n,out}\right]\right]}{\theta t_{\scriptscriptstyle slot}}.
\label{Eq:ServiceEnvelopeParameters}
\end{equation}

\subsection{Delay Bound Estimation}\label{sec:DelayBoundEstimation}
Using the results from Sections \ref{sec:uRLLCTrafficModel} and \ref{sec:ServiceModel}, we can define $W_{\scriptscriptstyle m}$ as a function of the free parameters $\theta$ and $\delta$, as written in Eq.~ \ref{eq:Definite_delay_bound}. To estimate $W_{\scriptscriptstyle m}$, we need to solve the following optimization problem.\\
\noindent \textbf{Problem}~\texttt{DELAY\_BOUND\_CALCULATION}:
\begin{alignat}{2}
&\underset{\theta,\delta}{\text{min}}  \quad && \frac{2 t_{\scriptscriptstyle slot} \left[ \text{ln} \left(\frac{\varepsilon_{\scriptscriptstyle m}}{2} \right) + \text{ln} \left(1-\text{exp} \left[\smallminus \theta\delta\right ] \right)\right] }{\text{ln} \left[ \sum_{n=0}^{
N_{\scriptscriptstyle add}} \frac{\pi_{\scriptscriptstyle m,n}}{T_{\scriptscriptstyle m,n}} \sum_{i=1}^{T_{\scriptscriptstyle m,n}} \text{exp} \left[ \smallminus \theta c_{\scriptscriptstyle m,i}^{\scriptscriptstyle n,out}\right] \right]  + \delta \theta t_{\scriptscriptstyle slot}} \label{eq:Definite_delay_bound}.\\
&\text{s.t.: } \quad && \theta,\delta,\rho_{\scriptscriptstyle A_{\scriptscriptstyle m}},\rho_{\scriptscriptstyle S_{\scriptscriptstyle m}} >0, \label{eq:ProblemFormulationCons1}\\
& \quad && \rho_{\scriptscriptstyle S_{\scriptscriptstyle m}} - \rho_{\scriptscriptstyle A_{\scriptscriptstyle m}} > 2\delta. \label{eq:ProblemFormulationCons4}
\end{alignat}
   


The previous problem is defined by a non-convex objective function over a non-convex region. It involves the existence of multiple local minimums, thus performing an exhaustive search to find the optimal solution is not computationally tractable. For such reason, we propose the heuristic approach described in Algorithm \ref{alg:OptimumDelayBound}. 
\begin{algorithm}[!t]
\SetAlgoLined
\small
\textbf{Inputs:} $\varepsilon_{\scriptscriptstyle m}$, $\vec{x}_{\scriptscriptstyle D_{\scriptscriptstyle m}}$, $\vec{x}_{\scriptscriptstyle C_{\scriptscriptstyle m},n}$\;
\textbf{Initialization:} $\theta_{\scriptscriptstyle z}=1$, $\delta_{\scriptscriptstyle z}$, $y$, $y_{\scriptscriptstyle z} =0$, $stop=False$\;
 \While{stop == False}{
 Set $\theta_{\scriptscriptstyle z}=\theta_{\scriptscriptstyle z} \Delta$\;
 Compute $\rho_{\scriptscriptstyle A_{\scriptscriptstyle m}}(\theta_{\scriptscriptstyle z})$, $\rho_{\scriptscriptstyle S_{\scriptscriptstyle m}}(\theta_z)$ [See Eqs. \eqref{Eq:ArrivalEnvelopeParameters} and \eqref{Eq:ServiceEnvelopeParameters}]\;
  \If{ $\rho_{\scriptscriptstyle S_{\scriptscriptstyle m}}(\theta_{\scriptscriptstyle z}) > \rho_{\scriptscriptstyle A_{\scriptscriptstyle m}}(\theta_{\scriptscriptstyle z})$}{
   Compute $\delta_{\scriptscriptstyle z} = \left(\rho_{\scriptscriptstyle S_{\scriptscriptstyle m}}(\theta_{\scriptscriptstyle z}) - \rho_{\scriptscriptstyle A_{\scriptscriptstyle m}}(\theta_{\scriptscriptstyle z})\right)/2$\;
   Compute $y_{\scriptscriptstyle z} = \theta_{\scriptscriptstyle z} \delta_{\scriptscriptstyle z}$\;
   \eIf{$y_{\scriptscriptstyle z} > y$ }{
    Set $y=y_{\scriptscriptstyle z}$, $\theta_{\scriptscriptstyle opt} = \theta_{\scriptscriptstyle z}$, $\delta_{\scriptscriptstyle opt} = \delta_{\scriptscriptstyle z}$\; 
    Compute $W_{\scriptscriptstyle m}(\theta_{\scriptscriptstyle opt},\delta_{\scriptscriptstyle opt})$. See Eq. \eqref{eq:Definite_delay_bound}\;
   }{
   $stop = True$\;
   }
   }
 }
 \textbf{return:} $\theta_{opt}$, $\delta_{opt}$, $W_m$\;
 \caption{Calculation Delay Bound $W_m$}
 \label{alg:OptimumDelayBound}
\end{algorithm}
This algorithm takes as inputs the target violation probability $\varepsilon_{\scriptscriptstyle m}$ for service $m$ and the sample vectors $\vec{x}_{\scriptscriptstyle D_{\scriptscriptstyle m}}$ and $\vec{x}_{\scriptscriptstyle C_{\scriptscriptstyle m},n}$, and iteratively searches the values of $\theta$ and $\delta$ which minimize Eq. \eqref{eq:Definite_delay_bound}, i.e., $\theta_{\scriptscriptstyle opt}$ and $\delta_{\scriptscriptstyle opt}$. We experimentally observed the rate parameter $\rho_{\scriptscriptstyle S_{\scriptscriptstyle m}}$, i.e., Eq.~\eqref{Eq:ServiceEnvelopeParameters}, increases when $\theta$ decreases. Note that a greater $\rho_{\scriptscriptstyle S_{\scriptscriptstyle m}}$ reduces the delay bound $W_{\scriptscriptstyle m}$ as Eq. \eqref{eq:Middle_delay_bound} shows. Additionally, the numerator of Eq. \eqref{eq:Definite_delay_bound} is monotonically increasing with the product $\theta\delta$. 
Based on that, Algorithm \ref{alg:OptimumDelayBound} first reduces the value of $\theta_{\scriptscriptstyle z}$, i.e., the candidate value of $\theta_{\scriptscriptstyle opt}$, in each iteration (step 4). Such reduction is performed by the parameter $\Delta \in (0,1)$. Refining $\Delta$ towards 1 enhances the granularity of locating $W_{\scriptscriptstyle m}$, albeit at the cost of requiring a larger number of iterations. Then, Algorithm \ref{alg:OptimumDelayBound} computes the rate parameters $\rho_{\scriptscriptstyle A_{\scriptscriptstyle m}}(\theta_{\scriptscriptstyle z})$ and $\rho_{\scriptscriptstyle S_{\scriptscriptstyle m}}(\theta_{\scriptscriptstyle z})$ (step 5). If  $\rho_{\scriptscriptstyle S_{\scriptscriptstyle m}}(\theta_{\scriptscriptstyle z}) > \rho_{\scriptscriptstyle A_{\scriptscriptstyle m}}(\theta_{\scriptscriptstyle z})$, i.e., from constraints \eqref{eq:ProblemFormulationCons1}-\eqref{eq:ProblemFormulationCons4}, the algorithm computes the candidate value of $\delta$, i.e., $\delta_{\scriptscriptstyle z}$ (step 7). If the product $\theta_{\scriptscriptstyle z}\delta_{\scriptscriptstyle z}$ (step 8) improves the product computed in the previous iteration, $W_{\scriptscriptstyle m}$ gets updated. In such a case, the algorithm sets $\theta_{\scriptscriptstyle opt}$ and $\delta_{\scriptscriptstyle opt}$ (step 10), computes $W_{\scriptscriptstyle m}$ (step 11), and tries another iteration to reduce $W_{\scriptscriptstyle m}$. When the product $\theta_{\scriptscriptstyle z}\delta_{\scriptscriptstyle z}$ does not improve with respect to the result of the previous iteration, the algorithm stops (step 13) and considers the latter as the optimal solution. 

\section{Control Loops of \name}\label{sec:SNC-basedOrchestrator}
In this section, we explain how \name performs the near-\gls{RT} and \gls{RT} control loops. Specifically, we first describe how the samples of the cell capacity are computed as well as the estimation of the probabilities $\pi_{\scriptscriptstyle m,n}$. Then, we explain how the \gls{SNC}-based Controller \emph{xApp} decides $N_{\scriptscriptstyle m}^{\scriptscriptstyle min}$ $\forall m \in \mathcal{M}$  for the near-\gls{RT} control loop. Finally, we describe how the \gls{RT} Controller \emph{dApp} mitigates the violation probability.

\subsection{Computation of the Cell Capacity Samples}\label{sec:TrafficEstimator}\label{sec:CellCapacityEstimator}
The Cell Capacity Estimator \emph{xApp} performs the following steps to obtain a sample $c_{\scriptscriptstyle m,i}^{\scriptscriptstyle n,out} \in \vec{x}_{\scriptscriptstyle C_{\scriptscriptstyle m},n}$.
\begin{enumerate}
    \item For each transmitted packet $j$, it considers (a) the packet size $l_{\scriptscriptstyle j}$, and (b) the amount of \glspl{RB} required to transmit it, i.e., $N_{\scriptscriptstyle pkt,j}$. Note that we are assuming that a unique \gls{MCS} value is adopted to transmit each packet. It means this \emph{xApp} can compute the number of bits transmitted per \gls{RB} as $c_{\scriptscriptstyle RB,j}=l_{\scriptscriptstyle j}/N_{\scriptscriptstyle pkt,j}$. Based on that, it defines a vector $\vec{x}_{\scriptscriptstyle pkt,j} = \{c_{\scriptscriptstyle RB,j},c_{\scriptscriptstyle RB,j}\, \hdots,c_{\scriptscriptstyle RB,j}\}$, where the element $c_{\scriptscriptstyle RB,j}$ is repeated $N_{\scriptscriptstyle pkt,j}$ times.
    \item Performing the previous step for all the transmitted packets, this \emph{xApp} obtains a set of vectors $\vec{x}_{\scriptscriptstyle pkt,j}$ $\forall j \in \mathcal{J}_{\scriptscriptstyle OBS}$, where $\mathcal{J}_{\scriptscriptstyle OBS}$ denotes the set of packets transmitted in the last $T_{\scriptscriptstyle OBS}$ \glspl{TTI}. Based on that, this \emph{xApp} defines $\vec{x}_{\scriptscriptstyle con}=\{\vec{x}_{\scriptscriptstyle pkt,1},\vec{x}_{\scriptscriptstyle pkt,2}\,\hdots \vec{x}_{\scriptscriptstyle pkt,|\mathcal{J}_{OBS}|}\}$ as a vector which concatenates each measured vector $\vec{x}_{\scriptscriptstyle pkt,j}$.
    \item Based on $\vec{x}_{\scriptscriptstyle con}$, this \emph{xApp} groups its samples in set of $N_{\scriptscriptstyle m,n}^{\scriptscriptstyle set}=n+N_{\scriptscriptstyle m}^{\scriptscriptstyle min}$ consecutive samples. In turn, each group of $N_{\scriptscriptstyle m,n}^{\scriptscriptstyle set}$ consecutive samples defines a vector $\vec{x}_{\scriptscriptstyle m,i}^{\scriptscriptstyle n,out}$, where $i \in [1,T_{\scriptscriptstyle m,n}]$ represents the i-th vector. We define $T_{\scriptscriptstyle m,n}$ as the total number of built vectors. Note we have one vector $\vec{x}_{\scriptscriptstyle m,i}^{\scriptscriptstyle n,out}$ per sample $c_{\scriptscriptstyle m,i}^{\scriptscriptstyle n,out}$, i.e., see Eq.~\eqref{eq:MGFAvailableCapacityforSlice}.
    \item Finally, this \emph{xApp} obtains the sample $c_{\scriptscriptstyle m,i}^{\scriptscriptstyle n,out}$ as the sum of all the elements of the vector $\vec{x}_{\scriptscriptstyle m,i}^{\scriptscriptstyle n,out}$, i.e., $c_{\scriptscriptstyle m,i}^{\scriptscriptstyle n,out}=\sum_{z=1}^{N_{\scriptscriptstyle m,n}^{\scriptscriptstyle set}} \vec{x}_{\scriptscriptstyle m,i}^{\scriptscriptstyle n,out}\{z\}$.
\end{enumerate}

\subsection{{Mixture Density Networks for estimating the RB utilization}}
\label{sec:NN_rb_util}
\begin{figure}[t!]
\centering
\includegraphics[width=\columnwidth]{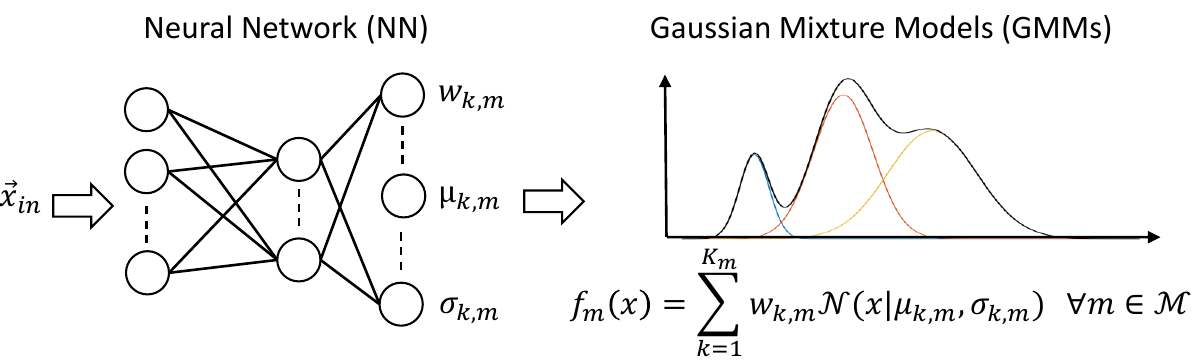}
\caption{Overview of the considered Mixture Density Network (MDN).}
\label{fig:MDN}
\end{figure}

The \gls{SNC}-based Controller \emph{xApp} relies on a \gls{MDN}, which is a \gls{NN} architecture designed to model probability distributions~\cite{MDN_book}, to estimate $\pi_{m,n}$. Typically, the \gls{NN}'s output layer produces a single value that represents the predicted outcome. In an \gls{MDN}, the output layer generates one or more parameterized mixture models, each being a weighted combination of several component distributions. In this paper, we consider $|\mathcal{M}|$ \glspl{GMM}, one per service. It is proven the \gls{GMM} accurately approximate any arbitrary distribution in the context of wireless networks~\cite{MDN_Modeling,Gaussian_Mixtures,GMM_basic}. Specifically, $\pi_{\scriptscriptstyle m,n}$ may not follow a single known statistical distribution and perhaps more importantly, it may also change over time. The \gls{GMM} is described by the equation depicted in Fig.~\ref{fig:MDN}, where $K_{\scriptscriptstyle m}$ denotes the number of Gaussian distributions. In turn, the $k$-th distribution is characterized by the weight $w_{\scriptscriptstyle k,m}$, the mean $\mu_{\scriptscriptstyle k,m}$ and the standard deviation $\sigma_{\scriptscriptstyle k,m}$. Note that $\sum_{k=1}^{K_{\scriptscriptstyle m}}w_{\scriptscriptstyle k,m}=1$.

The considered \gls{MDN} is summarized in Fig.~\ref{fig:MDN}. The inputs parameters $\vec{x}_{\scriptscriptstyle in}$ are: (a) the \gls{RB} utilization for each service, and (b) the 25th, 50th and 75th percentiles of incoming and enqueued bits. All of them measured in the last $T_{\scriptscriptstyle OUT}$ \glspl{TTI} for each service $m\in \mathcal{M}$. Additionally, the \gls{MDN} considers as input the target number of guaranteed \glspl{RB} for each service $m\in \mathcal{M}$ in the next $T_{\scriptscriptstyle OUT}$ \glspl{TTI}. Based on them, the \gls{MDN} provides an estimation of the parameters $w_{\scriptscriptstyle k,m}$, $\mu_{\scriptscriptstyle k,m}$ and $\sigma_{k,m}$. Finally, we compute the $\pi_{\scriptscriptstyle m,n}$ as Eq. \ref{eq:ProbabilitiesSNCCapacity} shows. Specifically, we split the \gls{GMM} into $N_{\scriptscriptstyle add}$ regions (i.e., see Section \ref{sec:ServiceModel}) and compute $\pi_{\scriptscriptstyle m,n} $  as the probability of being in the $n$-th region.
\begin{equation}
    \pi_{\scriptscriptstyle m,n} = \sum_{k=1}^{K_{\scriptscriptstyle m}}\int_{n-0.5}^{n+0.5} w_{\scriptscriptstyle k,m}\mathcal{N}\left(x|\mu_{\scriptscriptstyle k,m},\sigma_{\scriptscriptstyle k,m}\right) dx.
    \label{eq:ProbabilitiesSNCCapacity}
\end{equation}

\subsection{Near-RT Control Loop}\label{sec:ControlLoopRBAllocation}
The \gls{SNC}-based Controller \emph{xApp} aims to determine $N_{\scriptscriptstyle m}^{\scriptscriptstyle min}$ $\forall m \in \mathcal{M}$ such as the delay bound $W_{\scriptscriptstyle m}$ is as close as possible (or even below) to the target delay budget $W_{\scriptscriptstyle m}^{th}$, given the target violation probability $\varepsilon_{\scriptscriptstyle m}$. To that end, we formulate the problem  \texttt{ORCHESTRATION\_URLLC\_SERVICES}. It consists of minimizing the maximum ratio $W_{\scriptscriptstyle m}/W_{\scriptscriptstyle m}^{\scriptscriptstyle th}$, i.e., considering $|\mathcal{M}|$ services, as Eq. \eqref{eq:OptProblem} shows. This problem is constrained by Eq. \eqref{eq:OptORCons1}, which ensures the sum of the amount of guaranteed \glspl{RB} of all the services must be equal or less than the available \glspl{RB}.\\
\noindent \textbf{Problem}~\texttt{ORCHESTRATION\_URLLC\_SERVICES}:
\begin{alignat}{2}
&\underset{N_{\scriptscriptstyle m}^{\scriptscriptstyle min}}{\text{min.}}  \quad &&  g(\vec{W}) = \text{max}\left\{\sfrac{W_{\scriptscriptstyle 1}}{W_{\scriptscriptstyle 1}^{\scriptscriptstyle th}} ,\hdots, \sfrac{W_{\scriptscriptstyle |\mathcal{M}|}}{W_{\scriptscriptstyle |\mathcal{M}|}^{\scriptscriptstyle th}}\right\}. \label{eq:OptProblem}\\
&\text{s.t.: } \quad && \sum_{m=1}^{|\mathcal{M}|} N_{\scriptscriptstyle m}^{\scriptscriptstyle min} \leq 
N_{\scriptscriptstyle cell}^{\scriptscriptstyle RB}. \label{eq:OptORCons1}
\end{alignat}

The objective function $g(\vec{W})$ depends on the computation of $W_{\scriptscriptstyle m}$ $\forall m \in \mathcal{M}$. Specifically, when a specific value of $N_{\scriptscriptstyle m}^{\scriptscriptstyle min}$ is considered for each service, $|\mathcal{M}|$ sub-problems as \texttt{DELAY\_BOUND\_CALCULATION}, i.e., see Eqs. \eqref{eq:Definite_delay_bound}-\eqref{eq:ProblemFormulationCons4}, must be solved. Since each sub-problem requires the optimization of a non-convex function over a non-convex region, we propose Algorithm \ref{alg:OptimumOrchestration} to solve this problem.

\begin{algorithm}[t!]
\SetAlgoLined
\small
\textbf{Inputs:} $W_{\scriptscriptstyle m}^{\scriptscriptstyle th}$, $\varepsilon_{\scriptscriptstyle m}$, $\vec{x}_{\scriptscriptstyle D_{\scriptscriptstyle m}}$, $\vec{x}_{\scriptscriptstyle C_{\scriptscriptstyle m},n}$\;
\textbf{Initialization:} $N_{\scriptscriptstyle m,z}^{\scriptscriptstyle min}=\lfloor 
N_{\scriptscriptstyle cell}^{\scriptscriptstyle RB}/|\mathcal{M}|\rfloor$, $W_{\scriptscriptstyle m}=\infty$, $W_{\scriptscriptstyle m,z}=\infty$ $stop=False$\;
 \While{stop == False}{
 Estimate $\pi_{\scriptscriptstyle m,n}$ $\forall m \in \mathcal{M}$ $\forall n \in [0,
N_{\scriptscriptstyle add}]$ [Section \ref{sec:NN_rb_util}]\;
 Estimate $W_{\scriptscriptstyle m,z}$ $\forall m \in \mathcal{M}$ [See Algorithm \ref{alg:OptimumDelayBound}]\;
 Evaluate $g(\vec{W}_{\scriptscriptstyle z})$ and $g(\vec{W})$ [See Eq. \eqref{eq:OptProblem}]\;
  \eIf{$g(\vec{W}_{\scriptscriptstyle z}) <  g(\vec{W})$ }{
  Update $N_{\scriptscriptstyle m}^{\scriptscriptstyle min} = N_{\scriptscriptstyle m,z}^{\scriptscriptstyle min} $;  $W_{\scriptscriptstyle m} = W_{\scriptscriptstyle m,z}$  $\forall m \in \mathcal{M}$\;
  Select $m^{\scriptscriptstyle \prime} | \sfrac{W_{\scriptscriptstyle m^{\scriptscriptstyle \prime}}}{W_{\scriptscriptstyle m^{\scriptscriptstyle \prime}}^{\scriptscriptstyle th}} \geq  \sfrac{W_{\scriptscriptstyle m}}{W_{\scriptscriptstyle m}^{\scriptscriptstyle th}}$ $\forall m \in \mathcal{M} \setminus \{m^{\scriptscriptstyle \prime}\}$\;
  Select $m^{\scriptscriptstyle \prime\prime} | \sfrac{W_{\scriptscriptstyle m^{\scriptscriptstyle \prime\prime}}}{W_{\scriptscriptstyle m^{\scriptscriptstyle \prime\prime}}^{\scriptscriptstyle th}} \leq  \sfrac{W_{\scriptscriptstyle m}}{W_{\scriptscriptstyle m}^{\scriptscriptstyle th}}$ $\forall m \in \mathcal{M} \setminus \{m^{\scriptscriptstyle \prime\prime}\}$\;
  Compute $N_{\scriptscriptstyle m^{\scriptscriptstyle \prime},z}^{min} = N_{\scriptscriptstyle m^{\scriptscriptstyle \prime}}^{min} + 1$; $N_{\scriptscriptstyle m^{\scriptscriptstyle \prime\prime},z}^{\scriptscriptstyle min} = N_{\scriptscriptstyle m^{\scriptscriptstyle \prime\prime}}^{min} - 1$\;
   }{
   $stop = True$\;
   }
 }
 \textbf{return:} $N_{m}^{min}$, $W_m$\;
 \caption{Near-RT RB allocation}
 \label{alg:OptimumOrchestration}
\end{algorithm}

This algorithm considers as inputs the target delay budget $W_{\scriptscriptstyle m}^{\scriptscriptstyle th}$, the target violation probability $\varepsilon_{\scriptscriptstyle m}$ and the sample vectors $\vec{x}_{\scriptscriptstyle D_{\scriptscriptstyle m}}$, $\vec{x}_{\scriptscriptstyle C_{\scriptscriptstyle m},n}$. Additionally, it considers as starting point an equal distribution of the available \glspl{RB} among the services, i.e.,  $N_{\scriptscriptstyle m,z}^{\scriptscriptstyle min}=\lfloor 
N_{\scriptscriptstyle cell}^{\scriptscriptstyle RB}/|\mathcal{M}|\rfloor$. Based on them, Algorithm \ref{alg:OptimumOrchestration} starts an iterative procedure to get $N_{\scriptscriptstyle m}^{\scriptscriptstyle min}$ $\forall m \in \mathcal{M}$. In each iteration, considering $N_{\scriptscriptstyle m,z}^{\scriptscriptstyle min}$ guaranteed \glspl{RB} for each service, it first estimates $\pi_{\scriptscriptstyle m,n}$ using the \gls{MDN} described in Section \ref{sec:NN_rb_util} (step 4). Then, it uses the \gls{SNC}-based model (see Section \ref{sec:DelayBoundEstimation}) to estimate the delay bound $W_{\scriptscriptstyle m,z}$ for the target $N_{\scriptscriptstyle m,z}^{\scriptscriptstyle min}$ (step 5). Then, it evaluates the objective function, i.e., Eq. \eqref{eq:OptProblem}, considering $W_{\scriptscriptstyle m,z}$ and $W_{\scriptscriptstyle m}$ (step 6). Based on that, it checks if the objective function has been reduced with respect to the previous iteration (step 7). If so, this algorithm updates the new values for $N_{\scriptscriptstyle m}^{\scriptscriptstyle min}$ and $W_{\scriptscriptstyle m}$ (step 8) and tries to reduce the objective function. To that end, it selects the service $m^{\scriptscriptstyle \prime}$ with the best ratio $W_{m^{\scriptscriptstyle \prime}}/W_{\scriptscriptstyle m^{\scriptscriptstyle \prime}}^{\scriptscriptstyle th}$ and the service $m^{\scriptscriptstyle \prime\prime}$ with the worst ratio $W_{m^{\scriptscriptstyle \prime\prime}}/W_{\scriptscriptstyle m^{\scriptscriptstyle \prime\prime}}^{\scriptscriptstyle th}$ (steps 9-10). Next, Algorithm \ref{alg:OptimumOrchestration} takes one guaranteed \gls{RB} from the service $m^{\scriptscriptstyle \prime\prime}$ to be assigned to the service $m^{\scriptscriptstyle \prime}$ (step 11). Finally, a new iteration of the algorithm starts if $N_{\scriptscriptstyle m,z}^{\scriptscriptstyle min}$ $\forall m \in \mathcal{M}$ improves the objective function. Conversely, the iterative procedure ends if the objective function can not be further improved (step 13).

\subsection{RT Control Loop}\label{sec:InnerControlLoop}
Every $T_{\scriptscriptstyle OUT}$ \glspl{TTI}, the \gls{RT} Controller \emph{dApp} receives the new value of $N_{\scriptscriptstyle m}^{\scriptscriptstyle min}$ $\forall m \in \mathcal{M}$ from the \gls{SNC}-based Controller \emph{xApp}. Based on them, the \gls{RT} Controller \emph{dApp} operates at each \gls{TTI} as follows. First, it tries to drain the transmission queue of each service $m$ by using $N_{\scriptscriptstyle m}^{\scriptscriptstyle min}$ \glspl{RB}. After, a subset $\mathcal{M}^{\scriptscriptstyle \prime} \subset \mathcal{M}$ of services will have drained their queues, while the remaining $\mathcal{M}^{\scriptscriptstyle \prime\prime}=\mathcal{M}-\mathcal{M}^{\scriptscriptstyle \prime}$  services may still have pending transmissions. Assuming the $|\mathcal{M}^{\scriptscriptstyle \prime}|$ services have not fully consumed $N_{\scriptscriptstyle fr}$ \glspl{RB}, such spare resources can be used to for the remaining $|\mathcal{M}^{\scriptscriptstyle \prime\prime}|$ services. Specifically, the $N_{\scriptscriptstyle fr}$ \glspl{RB} will be allocated among the $|\mathcal{M}^{\scriptscriptstyle \prime\prime}|$ services following the \gls{EDF} discipline~\cite{EDF_scheduler} since it minimizes the number of packets whose transmission delay is above the target delay budget. Note that \gls{EDF} does not consider the violation probability~\cite{Elgabli2019}. For this reason, the \gls{RT} Controller \emph{dApp} combines \gls{EDF} with the establishment of guaranteed \glspl{RB} per service. Since the latter are decided by the \gls{SNC}-based Controller \emph{xApp}, our framework ensures $\text{P}\left[w>W_{\scriptscriptstyle m}\right]\leq \varepsilon_{\scriptscriptstyle m}$ $\forall m \in \mathcal{M}$ as long as the traffic and channel conditions do not change with respect to the samples $\vec{x}_{\scriptscriptstyle D_{\scriptscriptstyle m}}$, $\vec{x}_{\scriptscriptstyle C_{\scriptscriptstyle m},n}$.

If the traffic and/or channel conditions change, the probability the packet transmission delay $w$ is greater than the estimated delay budget $W_{\scriptscriptstyle m}$ may be greater than the violation probability, i.e., $\text{P}\left[w>W_{\scriptscriptstyle m}\right]\geq \varepsilon_{\scriptscriptstyle m}$. To avoid it whenever possible, the \gls{RT} Controller \emph{dApp} executes Algorithm \ref{alg:RT_operation}.

\begin{algorithm}[t!]
\SetAlgoLined
\small
\textbf{Inputs:} $N_{\scriptscriptstyle m}^{\scriptscriptstyle min}$, $Q_{\scriptscriptstyle T,m}^{\scriptscriptstyle U}$, $Q_{\scriptscriptstyle T,m}^{\scriptscriptstyle L}$, $\vec{s}_{\scriptscriptstyle i-1}$\;
Compute $q_{\scriptscriptstyle i,m}$ $\forall m \in \mathcal{M}$\;
Update states $\vec{s}_{\scriptscriptstyle i}$ and $N_{\scriptscriptstyle m,i}^{\scriptscriptstyle req}$ according to Fig. \ref{fig:STDtransmissionQueue}\;

$\vec{v}_{\scriptscriptstyle d}=\big\{m^{\scriptscriptstyle \prime}\big\}$ $\forall m^{\scriptscriptstyle \prime}$ $|\vec{s}_{\scriptscriptstyle i}\{m^{\scriptscriptstyle \prime}\}=A$ \;
$\vec{v}_{\scriptscriptstyle b}=\big\{m^{\scriptscriptstyle \prime\prime}\big\}$  $\forall m^{\scriptscriptstyle \prime\prime}$ $|\vec{s}_{\scriptscriptstyle i}\{m^{\scriptscriptstyle \prime\prime}\}=B\cup C$\;
\If{$\vec{v}_d \neq \emptyset$}{
Set $N_{\scriptscriptstyle ite}=\sum_{m^{\scriptscriptstyle \prime\prime}} N_{m^{\scriptscriptstyle \prime\prime},i}^{\scriptscriptstyle req}$ and $N_{\scriptscriptstyle m,i}^{\scriptscriptstyle min} = N_{\scriptscriptstyle m}^{\scriptscriptstyle min}$ $\forall m \in \mathcal{M}$\;
Set $j_{\scriptscriptstyle d} = 0$ and $j_{\scriptscriptstyle b}=0$\;
\For{$u$ from 1 to $N_{ite}$}{
Determine $n^{\scriptscriptstyle \prime} =\vec{v_{\scriptscriptstyle d}}\{j\}$ and $n^{\scriptscriptstyle \prime\prime} = \vec{v_{\scriptscriptstyle b}}\{j\}$\;
Update $N_{\scriptscriptstyle n^{\scriptscriptstyle \prime},i}^{\scriptscriptstyle min} = N_{\scriptscriptstyle n^{\scriptscriptstyle \prime},i}^{\scriptscriptstyle min} - 1$; $N_{\scriptscriptstyle n^{\scriptscriptstyle \prime\prime},i}^{\scriptscriptstyle min} = N_{\scriptscriptstyle n^{\scriptscriptstyle \prime\prime},i}^{\scriptscriptstyle min} + 1$\;
Update $j_{\scriptscriptstyle d} = j_{\scriptscriptstyle d} + 1$ and $j_{\scriptscriptstyle b} = j_{\scriptscriptstyle b} + 1$\;
\If{$j_{\scriptscriptstyle d} == |\vec{v}_{\scriptscriptstyle d}|$}{
$j_{\scriptscriptstyle d} = 0$
}
\If{$j_{\scriptscriptstyle b} == |\vec{v}_{\scriptscriptstyle b}|$}{
$j_{\scriptscriptstyle b} = 0$
}
}

}
 \textbf{return:} $N_{\scriptscriptstyle m,i}^{\scriptscriptstyle min}$, $N_{\scriptscriptstyle m,i}^{\scriptscriptstyle req}$\;
 \caption{Mitigating $w> W_{\scriptscriptstyle m}^{\scriptscriptstyle th}$ at TTI $i$}
 \label{alg:RT_operation}
\end{algorithm}

\begin{figure}[t!]
    \centering
    \includegraphics[width=\columnwidth]{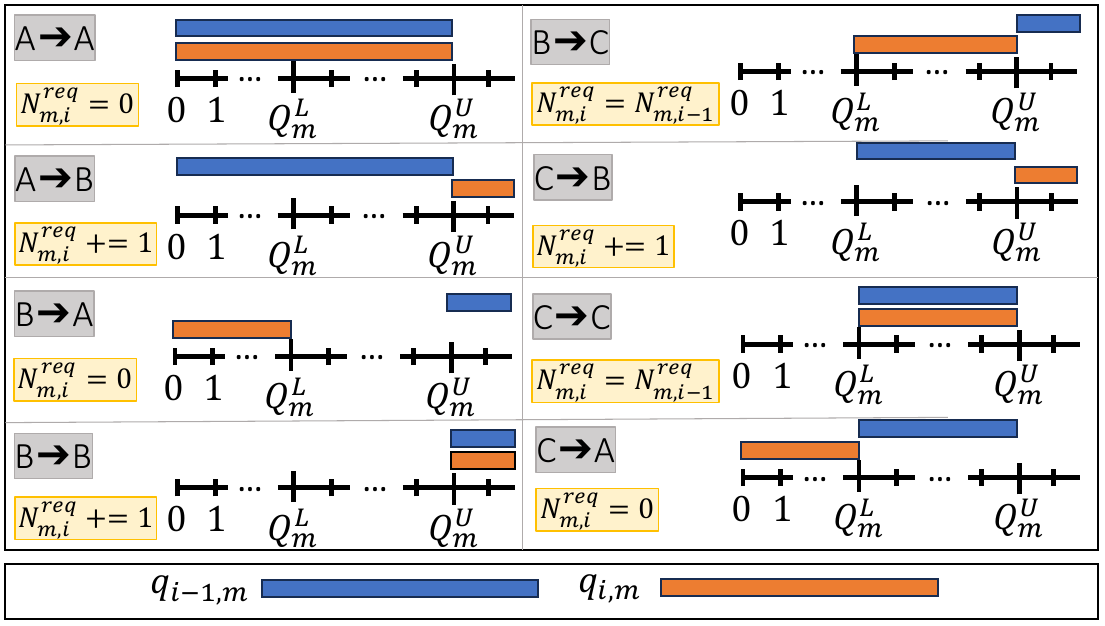}
    \caption{Transitions of the finite-state machine to control the \gls{RB} allocation for service $m$.}
    \label{fig:STDtransmissionQueue}
\end{figure}

This algorithm monitors the waiting time of the first packet of each service in the transmission queue. Then, if the waiting time is close to the delay budget, the algorithm increases (if possible) the amount of guaranteed \glspl{RB} for the corresponding service. To this end, Algorithm \ref{alg:RT_operation} relies on a finite-state machine of three states $\big\{A,B,C\big\}$ based on two thresholds $Q_{\scriptscriptstyle T,m}^{\scriptscriptstyle U}$ and $Q_{\scriptscriptstyle T,m}^{\scriptscriptstyle L}$. The threshold $Q_{\scriptscriptstyle T,m}^{\scriptscriptstyle U}=\eta Q_{\scriptscriptstyle T,m}$ indicates the waiting time of a packet is close to the delay budget, whereas $Q_{\scriptscriptstyle T,m}^{\scriptscriptstyle L}= \tau Q_{\scriptscriptstyle T,m}$ indicates the waiting time is far to the delay budget. The parameter $Q_{\scriptscriptstyle T,m}=\lfloor W_{\scriptscriptstyle m}^{\scriptscriptstyle th} / t_{\scriptscriptstyle slot} \rfloor$ is the maximum number of \glspl{TTI} that a packet can wait in the queue before crossing the delay budget. Additionally, $Q_{\scriptscriptstyle T,m}^{\scriptscriptstyle U} > Q_{\scriptscriptstyle T,m}^{\scriptscriptstyle L}$. Note that $\eta \in (0,1]$ and $\tau \in (0,1]$ can be tuned by the \gls{MNO}. Regarding the states, the state $A$ indicates the \gls{RT} Controller \emph{dApp} allocates to the service $m$ the amount of guaranteed \glspl{RB} decided by the \gls{SNC}-based Controller \emph{dApp}, i.e., $N_{\scriptscriptstyle m,i}^{\scriptscriptstyle min} = N_{\scriptscriptstyle m}^{\scriptscriptstyle min}$. Note that we define $N_{\scriptscriptstyle m,i}^{\scriptscriptstyle min}$ as the amount of guaranteed \glspl{RB} decided by the \gls{RT} Controller \emph{dApp} in the \gls{TTI} $i$. The state $B$ indicates the waiting time of the first packet of service $m$ is very close to the delay budget, thus the \gls{RT} Controller \emph{dApp} may increase the amount of guaranteed \glspl{RB} for such service. Specifically, it may increase $N_{\scriptscriptstyle m,i}^{\scriptscriptstyle req}$ \glspl{RB}. In state $B$, $N_{m,i}^{\scriptscriptstyle req}$ increases by one \gls{RB} with respect to the previous \gls{TTI}. The state $C$ indicates the waiting time of the first packet is lower than in state $B$, but not enough to go back to  $N_{\scriptscriptstyle m,i}^{\scriptscriptstyle min} = N_{\scriptscriptstyle m}^{\scriptscriptstyle min}$. In such case, Algorithm \ref{alg:RT_operation} keeps the same value of $N_{\scriptscriptstyle m,i}^{\scriptscriptstyle req}$ with respect to the previous \gls{TTI}. In Fig.~\ref{fig:STDtransmissionQueue} we summarize the possible transitions among states. Considering such transitions, we define $\vec{s}_{\scriptscriptstyle i}$ as a vector containing the state for each service at \gls{TTI} $i$.

Based on $N_{\scriptscriptstyle m}^{\scriptscriptstyle min}$, $Q_{\scriptscriptstyle T,m}^{\scriptscriptstyle L}$, $Q_{\scriptscriptstyle T,m}^{\scriptscriptstyle U}$ and $\vec{s}_{\scriptscriptstyle i-1}$, Algorithm \ref{alg:RT_operation} initially computes the state of the first packet of each service as $q_{\scriptscriptstyle i,m}=w_{\scriptscriptstyle m}^{\scriptscriptstyle pkt}/t_{\scriptscriptstyle slot}$ (step 2). Note that $w_{\scriptscriptstyle m}^{\scriptscriptstyle pkt}$ is the waiting time of such a packet. Then, it updates the states $\vec{s}_{\scriptscriptstyle i}$ and $N_{\scriptscriptstyle m,i}^{\scriptscriptstyle req}$ according to the transitions depicted in Fig. \ref{fig:STDtransmissionQueue} (step 3). Later,  Algorithm \ref{alg:RT_operation} needs to check if the amount of \glspl{RB} defined in $N_{\scriptscriptstyle m,i}^{\scriptscriptstyle req}$ can be allocated, in addition to $N_{\scriptscriptstyle m}^{\scriptscriptstyle min}$, to the corresponding services. The policy considered by Algorithm~\ref{alg:RT_operation} is that only the services whose state is $A$ can donate \glspl{RB} to those which require more \glspl{RB}. Considering this policy, Algorithm \ref{alg:RT_operation} iteratively re-allocates the amount of guarantees \glspl{RB} from services in state $A$ to services in states $B$ or $C$ (steps 4-20).  

\section{Performance Evaluation}\label{sec:Results}
\begin{figure}[b!]
    \centering
    \includegraphics[width=\columnwidth]{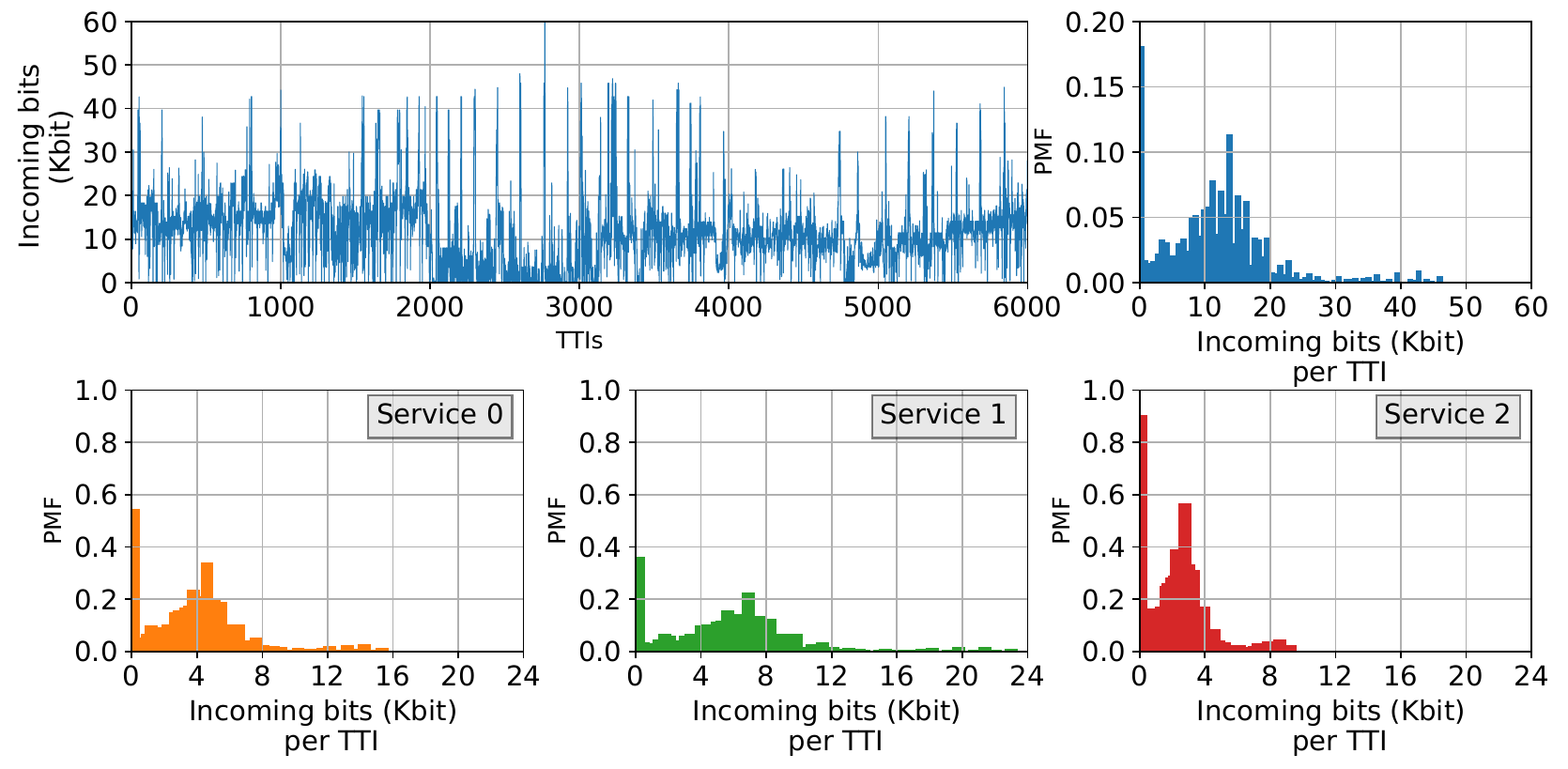}
   \caption{PMFs of incoming bits from FALCON traces.}
    \label{fig:InputServices}
\end{figure}

We perform an exhaustive simulation campaign to validate \name, and evaluate its performances through a Python-based simulator running on a computing platform with $16$ GB RAM and a quad-core Intel Core i7-7700HQ @ 2.80 GHz. The simulator considers a single cell using an \gls{OFDMA} scheme with $
N_{\scriptscriptstyle cell}^{\scriptscriptstyle RB}\in [50,100]$ \glspl{RB} and $t_{\scriptscriptstyle slot}$ = 1 ms. Regarding the incoming traffic and cell capacity of each \gls{uRLLC} service, we consider realistic traces collected over an operational RAN using the FALCON tool~\cite{falcon}. The tool allows decoding the Physical Downlink Control Channel (PDCCH) of a base station, revealing the number of active users and their scheduled resources. We make the traces public to foster research on the topic and favor reproducibility\footnote{FALCON traces. Online available: \url{https://nextcloud.neclab.eu/index.php/s/tTqCfRHbgx8Xwtj}. Password: ORANUS\_INFOCOM24}. Fig.~\ref{fig:InputServices} depicts the incoming bits per \gls{TTI} measured by FALCON, as well as the corresponding \gls{PMF} (i.e., upper plots). To emulate the incoming traffic of three different \gls{uRLLC} services, we order the active \glspl{UE} and split them into equally-sized groups, considering their aggregated traffic demand. The resulting \glspl{PMF} are also displayed in Fig.~\ref{fig:InputServices} (lower plots).

For these services, we set a delay budget $W_{\scriptscriptstyle m}^{\scriptscriptstyle th}=\big\{5,10,15\big\}$ ms and target violation probability $\varepsilon_{\scriptscriptstyle m}=\big\{10^{-5},10^{-4},10^{-3}\big\}$ \cite{Haque2023,Siddiqui2023,Khan2022}.

\subsection{Validation of SNC model}
In the first experiment, we consider a cell hosting a single service $m$ with the incoming traffic measured by FALCON (i.e., blue plots in Fig.~\ref{fig:InputServices}). Assuming a target violation probability $\varepsilon_{\scriptscriptstyle m}=10^{-3}$, we compute the delay bound $W_{\scriptscriptstyle m}^{\scriptscriptstyle mod}$ using the proposed \gls{SNC} model and compare it with the real bound  $W_{\scriptscriptstyle m}^{\scriptscriptstyle sim}$ obtained by simulation while the \gls{SNC}-based Controller \emph{xApp} (a) allocates $N_{\scriptscriptstyle m}^{\scriptscriptstyle min} \in \left[50,100\right]$ \glspl{RB} for such service, and (b) uses $T_{\scriptscriptstyle OBS} \in \big\{1, 2, 3, 4, 5, 6 \big\}\cdot 1000$ \glspl{TTI} to obtain $\vec{x}_{\scriptscriptstyle D_m}$ and $\vec{x}_{\scriptscriptstyle C_{\scriptscriptstyle m},n}$. 

\begin{figure}[t!]
    \centering
    \includegraphics[width=\columnwidth]{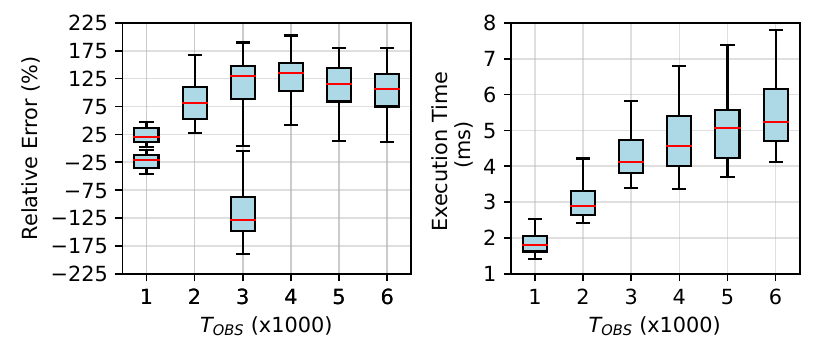}
    \vspace{-7mm}
   \caption{Validation SNC-based model.}
    \label{fig:Validation}
\end{figure}

The results are depicted in Fig.~\ref{fig:Validation}. On the left side, the box-and-whisker plot represents the distribution of the relative error $\epsilon_{\scriptscriptstyle r} = (W_{\scriptscriptstyle m}^{\scriptscriptstyle mod}-W_{\scriptscriptstyle m}^{\scriptscriptstyle sim})/W_{\scriptscriptstyle m}^{\scriptscriptstyle sim}\cdot 100$ when considering a variable $T_{\scriptscriptstyle OBS}$. Note that we have used two boxes in those scenarios which present negative relative errors. Specifically, each box gathers either the positive relative errors or the negative relative errors. We can notice the mean relative error is always below 150\%. Such value may appear large at first, but is in fact acceptable, as the goal of our \gls{SNC}-based model is to accommodate complex arrival and service processes at the expense of obtaining an exact match between the model and simulator results. Instead, \gls{SNC} promises an upper and conservative estimation of $W_{\scriptscriptstyle m}$, which is of key importance to meet uRLLC services' requirements in real scenarios.

From the same picture, we can observe negative values for $\epsilon_{\scriptscriptstyle r}$ when $T_{\scriptscriptstyle OBS}\leq 3000$ \glspl{TTI}. This is due to the insufficient number of samples in vectors $\vec{x}_{\scriptscriptstyle D_{\scriptscriptstyle m}}$ and $\vec{x}_{\scriptscriptstyle C_{\scriptscriptstyle m},n}$, which are not enough to capture the \glspl{PMF} for the incoming traffic and cell capacity for the corresponding service. In our settings, the proposed \gls{SNC} model needs at least a period of $T_{\scriptscriptstyle OBS} = 4000$ \glspl{TTI} to effectively capture such \glspl{PMF} and provide a meaningful upper estimation of the delay bound. 

On the right-hand side, the plot shows the execution time distribution of the proposed \gls{SNC} model. The time monotonically increases with $T_{\scriptscriptstyle OBS}$ as more samples need to be considered in Eqs. \eqref{Eq:ArrivalEnvelopeParameters}, \eqref{Eq:ServiceEnvelopeParameters} and \eqref{eq:Definite_delay_bound}. For the considered scenarios, the average execution time is below 6 ms, making the proposed model suitable for determining the amount of guaranteed \glspl{RB} for \gls{uRLLC} services in a near-\gls{RT} scale. 

\subsection{Validation of \gls{MDN}}
In Section~\ref{sec:NN_rb_util}, we propose a neural network model to expedite the estimation of $\pi_{m,n}$. To validate the proposed \gls{MDN} approach, we split the FALCON dataset according to a 70/30 ratio for the purposes of training and testing, respectively. Furthermore, the number of active \glspl{UE} are grouped in a uniform way to form the traffic demand of each service. In our implementation, the \gls{MDN} accounts for a $4$-staged network, characterized by $256$, $256$, $64$, and $3 \cdot K_{\scriptscriptstyle m} \cdot |\mathcal{M}|$ neurons, respectively. We empirically select the Rectified Linear Units (ReLu) as activation function to overcome the vanishing gradient problem and allow the model to learn faster. The kernel weights of each layer are initialized exploiting the He\_Normal statistical distribution. We also adopted kernel regularization techniques to regularize the learning and improve the generalization of the results. Finally, we adopt the \gls{MSE} metric to train the model and choose the Adam optimizer with a learning rate $0.001$ to optimize the loss function. 

\begin{figure}[t!]
\centering
\includegraphics[width=\columnwidth]{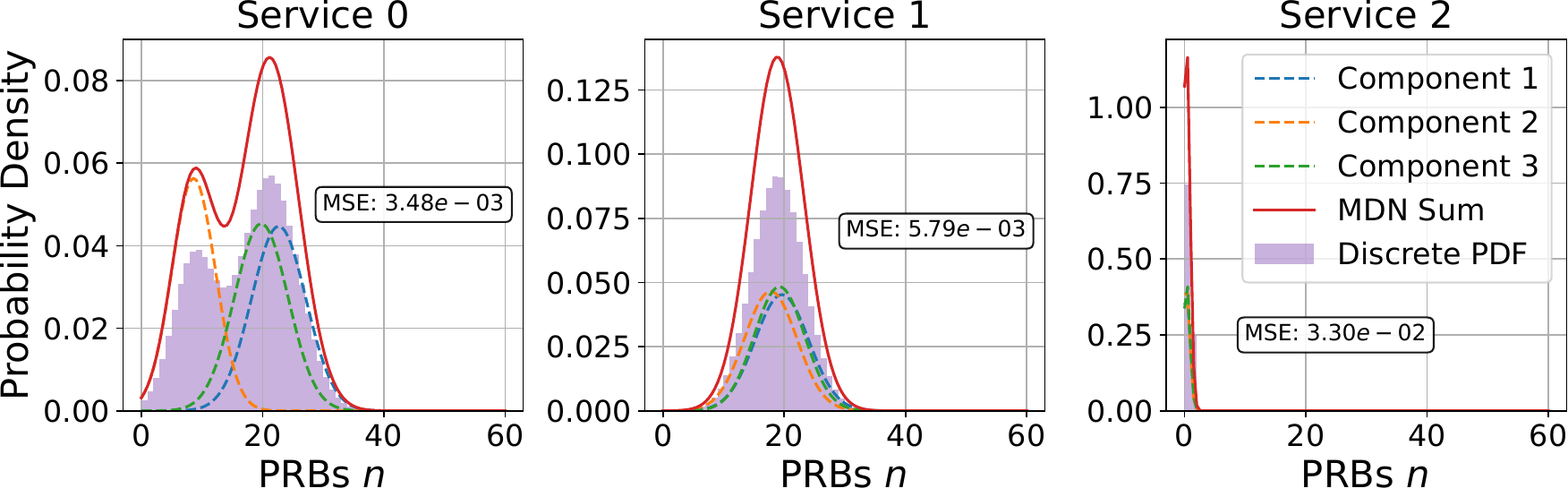}
\caption{Validation of the proposed MDN model.}
\label{fig:MDN_analysis}%
\end{figure}

The model is trained with multivariate input including incoming traffic demand, transmission queue size, and currently guaranteed PRBs over a monitoring time window of $T_{\scriptscriptstyle OBS} = 4000$ TTIs, aiming to estimate $w_{\scriptscriptstyle k,m}$, $\mu_{\scriptscriptstyle k,m}$ and  $\sigma_{\scriptscriptstyle k,m}$ $\forall k \in [0,K_{\scriptscriptstyle m}]$ $\forall m \in \mathcal{M}$. Fig.~\ref{fig:MDN_analysis} depicts the resulting PDF estimation assuming $K_{\scriptscriptstyle m}=3$. It can be noticed how the \gls{MDN} is able to estimate the continuous PDF distribution of the expected available PRBs per service even in the presence of heterogeneous shapes. Such curves are then discretized as Eq.~\eqref{eq:ProbabilitiesSNCCapacity} shows to obtain the \gls{PMF} $ \pi_{\scriptscriptstyle m,n}$ $\forall m \in \mathcal{M}$. 


\subsection{Performance Analysis of the SNC-based Controller xApp}
In a third set of experiments, we focus on a single decision period of the \gls{SNC}-based Controller \emph{xApp}. Specifically, we assume it computes $N_{\scriptscriptstyle m}^{\scriptscriptstyle min}$ for the three services described at the beginning of Section \ref{sec:Results}. Under this scenario, we evaluate the convergence of the heuristics proposed in Algorithm \ref{alg:OptimumOrchestration}, the computational complexity of such heuristics, and the accuracy of the obtained solution with respect to the optimal.
\begin{figure}[b!]
    \centering
    \includegraphics[width=\columnwidth]{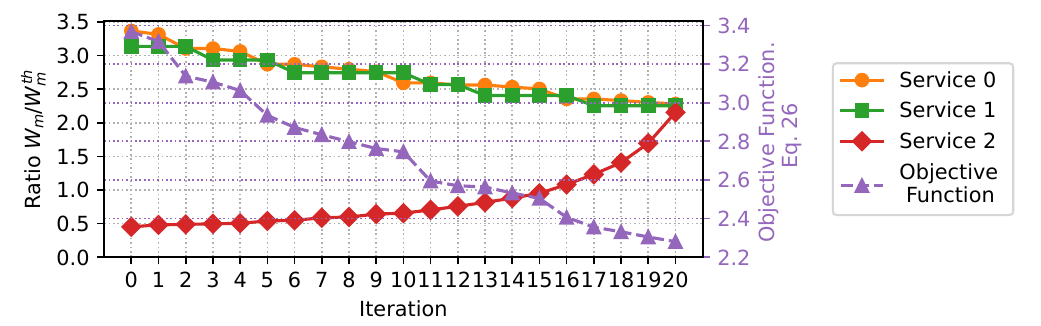}
    \caption{Convergence Analysis of Algorithm 
    \ref{alg:OptimumOrchestration}.}
    \label{fig:Orchestrator}
\end{figure}

\begin{figure}[t!]
    \centering
    \includegraphics[width=\columnwidth, clip, trim = 0cm 0.2cm 0cm 0cm]{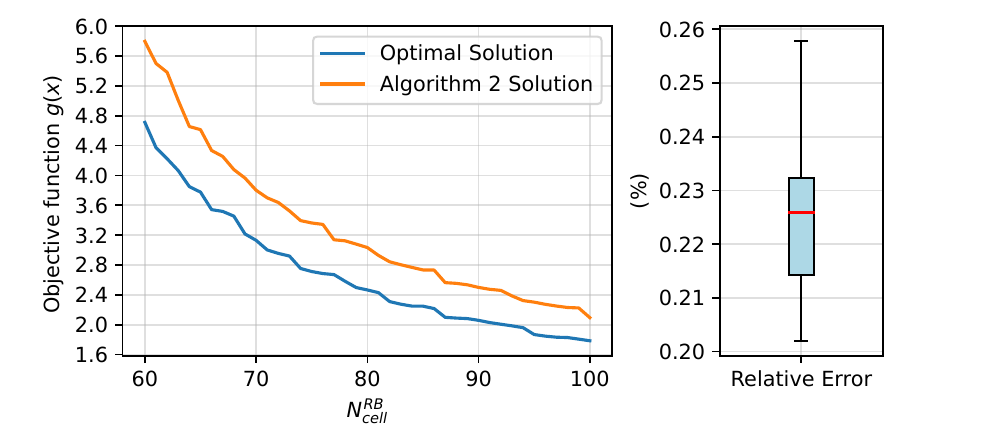}
    \caption{Performance Analysis of the \gls{SNC}-based Controller.}
   \vspace{-2mm}
   \label{fig:CompBruteForce}
\end{figure}

Fig.~\ref{fig:Orchestrator} shows the convergence of Algorithm \ref{alg:OptimumOrchestration}. Specifically, it depicts the value of the objective function $g(\vec{W})$, i.e., purple curve, as well as the values of the ratios $W_{\scriptscriptstyle m}/W_{\scriptscriptstyle m}^{\scriptscriptstyle th}$ $\forall m \in \mathcal{M}$, i.e., orange, green and red curves, in each iteration. We observe how the proposed heuristics iteratively reduces $g(\vec{W})$ until reaching a suboptimal solution. In Fig.~ \ref{fig:CompBruteForce}, we compare the solution obtained by the proposed heuristics with respect to the optimal one derived by brute force approach when considering $N_{\scriptscriptstyle cell}^{\scriptscriptstyle RB} \in [50,100]$ \glspl{RB}. The relative error between the two curves is approximately $0.225 \%$, indicating that the proposed solution yields only a 0.225\% increase in the worst ratio $\sfrac{W_{\scriptscriptstyle m}}{W_{\scriptscriptstyle m}^{\scriptscriptstyle th}}$ (see Eq. \eqref{eq:OptProblem}) compared to the brute force approach. It is a reasonable deviation given the significant differences in computational complexity among the two approaches as Table~\ref{tab:BruteForceComparison} shows. Specifically, Table~\ref{tab:BruteForceComparison}  summarizes the average execution time, the number of iterations, and the resulting ratio as a function of $N_{\scriptscriptstyle cell}^{\scriptscriptstyle RB}$.
We observe the number of iterations monotonically increases with $N_{\scriptscriptstyle cell}^{\scriptscriptstyle RB}$ as the search space (i.e., combinations of RB allocation for each service) is greater, whereas the average execution time for a single iteration slightly decreases. Note that the tasks performed in a single iteration are the same for both approaches.
The execution time per iteration decreases for larger $
N_{\scriptscriptstyle cell}^{\scriptscriptstyle RB}$ values. This is due to the fact that the estimation of $W_{\scriptscriptstyle m}$ (i.e., when Algorithm \ref{alg:OptimumOrchestration} calls Algorithm \ref{alg:OptimumDelayBound} in step 5) is faster if more \glspl{RB} are considered for each service. Specifically, we have experimentally observed that less iterations are needed by Algorithm \ref{alg:OptimumDelayBound} when $N_{\scriptscriptstyle cell}^{\scriptscriptstyle RB}$ increases, i.e., the value of $\theta_{\scriptscriptstyle z}$ (step 4) which minimizes Eq. \eqref{eq:Definite_delay_bound} is greater. 
%
\begin{table}[t!]
\centering
\caption{Computational Complexity Comparison}
\label{tab:BruteForceComparison}
\resizebox{\columnwidth}{!}{
\begin{tabular}{|c|c|c|c|c|c|}
\hline
\rule{0pt}{1.2em} 
\textbf{$\mathbf{N_{cell}^{RB}}$}                                             & \textbf{60} & \textbf{70} & \textbf{80} & \textbf{90} & \textbf{100} \\[4pt] \hline
\textbf{\begin{tabular}[c]{@{}c@{}}Brute Force\\ (iterations)\end{tabular}}  & 1711        & 2346           & 3081        & 3916        & 4851         \\ \hline
\textbf{\begin{tabular}[c]{@{}c@{}}Algorithm 2\\ (iterations)\end{tabular}} & 12           & 14          & 16                  & 19                & 22           \\ \hline
\textbf{\begin{tabular}[c]{@{}c@{}} Ratio Algorithm 2\\ / Brute Force \end{tabular}} & 142.58           & 167.57          & 192.56                  & 206.10                & 220.5           \\ \hline
\textbf{\begin{tabular}[c]{@{}c@{}} Avg. Execution Time per \\ iteration (ms)\end{tabular}} & 184.62             & 179.09            & 176.39               & 173.76               & 171.85           \\ \hline
\end{tabular}
}
\end{table}

\subsection{Performance Analysis of \name framework}
In the last experiment, we evaluated the performance of \name against three reference solutions. The reference solution \#1 consists of a single \gls{RT} Controller using the \gls{EDF} discipline. The reference solution \#2 only considers the \gls{SNC}-based Controller \emph{xApp}. Note that this solution allocates dedicated \glspl{RB} per service without sharing. The reference solution \#3 considers the \gls{SNC}-based Controller \emph{xApp} and the \gls{RT} Controller \emph{dApp}. However, the \gls{RT} Controller \emph{dApp} does not use Algorithm~\ref{alg:RT_operation}. 

\begin{figure}[t!]
    \centering
    \includegraphics[width=\columnwidth]{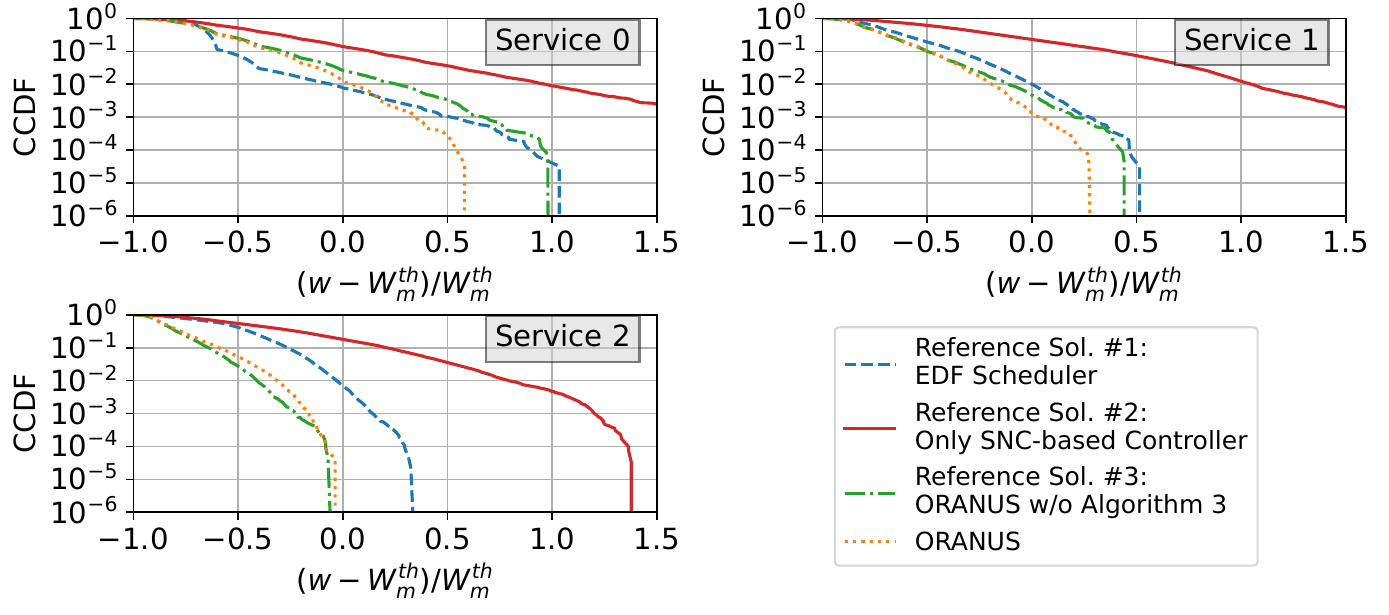}
   \caption{Complementary Cumulative Distribution Function (CCDF) for \\ $(w-W_{\scriptscriptstyle m}^{\scriptscriptstyle th})/W_{\scriptscriptstyle m}^{\scriptscriptstyle th}$. In \name, we have set  $\eta = 0.75$ and $\tau = 0.3$.}
   \label{fig:SLA_analysis}
\end{figure}

To measure the performance of these solutions, we consider the \gls{CCDF} of the metric $(w-W_{\scriptscriptstyle m}^{\scriptscriptstyle th})/W_{\scriptscriptstyle m}^{\scriptscriptstyle th}$. Note the random variable $w$ represents the transmission delay of an arbitrary packet. Additionally, when this metric is equal to 0, the \gls{CCDF} value represents the violation probability. In Fig.~\ref{fig:SLA_analysis} we observe that the reference solution \#2 provides the worst performance, i.e., a violation probability $11.46$ and $178.30$ times larger than \name for services 1 and 2. Note that this probability is equal to 0 for scenario 3 when we consider \name. These results are due to the fact the reference solution \#2 only considers dedicated \glspl{RB}. The remaining solutions consider sharing \glspl{RB} among the services. Comparing them, the reference solution \#1 usually provides a greater violation probability with respect to \name. Although \gls{EDF} ensures the packet with the earliest deadline are transmitted first, e.g., we observe for service 0 the \gls{CCDF} is lower for reference solution \#1 when $(w-W_{\scriptscriptstyle m}^{\scriptscriptstyle th})/W_{\scriptscriptstyle m}^{\scriptscriptstyle th} \leq 0$, it does not provide any guarantees in terms of violation probability. To consider such probability, the \name framework establishes guaranteed \glspl{RB} in a near-\gls{RT} and allocates share free \glspl{RB} among services using \gls{EDF} in a \gls{RT} scale. It improves the performance with respect to the remaining solutions. Specifically, \name provides lower violation probabilities for service \#1 and service \#2. Finally, we observe the consideration of Algorithm \ref{fig:STDtransmissionQueue} in \name improves the obtained violation probability if we compare the results with respect to the ones obtained by the reference solution \#3. The improvement is most significant when the metric $(w-W_{\scriptscriptstyle m}^{\scriptscriptstyle th})/W_{\scriptscriptstyle m}^{\scriptscriptstyle th}$ is higher. This is due to Algorithm \ref{fig:STDtransmissionQueue} performing its actions when the waiting time of a packet in the transmission queue is closer to the delay budget. We can also observe the reference solution \#3 has a similar behavior as \name for service 2. The reason is the packet transmission delay never is above the delay budget for this service, as happens for services 0 and 1.

\section{Related Work}\label{sec:RelatedWorks}
Several works addressed the radio resource allocation problem in the presence of multiple \gls{uRLLC} services. Concerning \gls{O-RAN} based solutions, the authors of~\cite{Abedin2022} propose a framework based on the actor-critic algorithm to minimize the probability of \gls{IoT} devices' age of information to exceed a predefined threshold. Despite its novelty, this work does not consider the transmission delay in the air interface. In~\cite{Karbalaee2023}, the authors propose an iterative algorithm to address the joint radio resource and power allocation problem, while the authors of~\cite{Rezazadeh2023} solve this problem by federated learning. In spite of their significance, their findings focus on the average delay as the key parameter. The authors of~\cite{PoleseColo} analyze the performance of \gls{DRL} agents using a non-\gls{RT} control loop that implements control actions in \gls{O-RAN}. Despite its novelty, this solution omits details on the interactions of multi-time-scale control loops.

Focusing on non-\gls{RT} and near-\gls{RT} solutions, some works rely on queuing theory to model the packet transmission delay~\cite{Zhang2023,Shi2022,Yang2021}. However, these models can only obtain average values when complex distributions are considered for the packet arrival rate and the channel capacity. Other works such as~\cite{SOTANC1,SOTANC2,SOTANC3} use \gls{SNC} to estimate a bound $W$ of type $\text{P}[w>W]<\varepsilon$, where $w$ is the packet transmission delay and $\varepsilon$ a target tolerance. However, they do not assume scenarios involving multiple \gls{uRLLC} services nor their cross-interference. In~\cite{Adamuz-Hinojosa-TWC2023}, the authors propose a \gls{SNC}-based controller for planning multiple \gls{uRLLC} services. However, it considers dedicated radio resources for each service, which may result in resource wastage. Additionally, it is limited to traffic arrivals that follow a Poisson process with batches.

Considering \gls{RT} solutions, works like~\cite{EDF_scheduler,Hadar2018,Raviv2023} use schedulers based on \gls{EDF} to assign priorities to packets based on their deadlines. \gls{EDF} ensures the packets with the earliest deadline are transmitted first. However, \gls{EDF} does not consider the probability the packet transmission delay exceeds a delay budget~\cite{Capozzi2013}. Other solutions such as~\cite{Zhang2019,Esswie2020,Alsenwi2021} rely on \gls{ML} models. Although they are effective in managing scenarios with intricate traffic patterns and channel conditions, their performance is primarily reliant on the similarity between the measured patterns and those used during training.

\gls{O-RAN} specifications mention a \gls{RT} control loop for optimizing tasks such as packet scheduling or interference recognition~\cite{O-RAN-WG2-AIML}. However, at the moment of writing this paper, such a control loop has not been defined. In the same row, the authors of \cite{dApps_article} introduce the concept of \emph{dApps} to implement fine-grained \gls{RT} control tasks. Despite implementing a proof-of-concept, they omit to detail how multiple \gls{uRLLC} services can be orchestrated in a \gls{RT} scale, and how the near-\gls{RT} control loop interacts with the \emph{dApps}.

\section{Conclusions}\label{sec:Conclusions}
In this paper, we addressed the need of effective multi-time-scale control loops in \gls{O-RAN}-based deployments for \gls{uRLLC} services. 
Specifically, we proposed \name, an O-RAN-compliant mathematical framework focused on the radio resource allocation problem at near-\gls{RT} and \gls{RT} scales. Focusing on the near-\gls{RT} control loop, \name relies on a novel \gls{SNC} model to compute the amount of guaranteed \glspl{RB} per service. Unlike traditional approaches as queueing theory, the \gls{SNC} model allows \name ensuring the probability the packet transmission delay exceeds a specific budget, i.e., the violation probability, is below a target tolerance. Another key novelty of \name is the incorporation of a \gls{RT} control loop which monitors the transmission queue of each service and dynamically adjusts the allocation of guaranteed \glspl{RB} in response to traffic anomalies. We evaluated our proposal by a comprehensive simulation campaign, where \name demonstrated substantial improvements, with an average violation probability $10\times$ lower, in comparison to reference solutions.


\newpage
\bibliographystyle{ieeetr}
\bibliography{references}

\end{document}

%% file: acronyms.tex
\newacronym{3GPP}{3GPP}{3rd Generation Partnership Project}

\newacronym{5G}{5G}{5th Generation}
\newacronym{5G-ACIA}{5G-ACIA}{5G Alliance for Connected Industries and Automation}
\newacronym{5G-NR}{5G-NR}{5G New Radio}
\newacronym{6G}{6G}{Sixth Generation}

\newacronym{AMC}{AMC}{Adaptive Modulation and Coding}
\newacronym{AC}{AC}{Admission Control}
\newacronym{AGV}{AGV}{Automated Guided Vehicle}
\newacronym{AR}{AR}{Augmented Reality}

\newacronym{BLER}{BLER}{Block Error Rate}
\newacronym{BWP}{BWP}{Bandwidth Part}
\newacronym{BSS}{BSS}{Business Support System}

\newacronym{CDF}{CDF}{Cumulative Distribution Function}
\newacronym{CCDF}{CCDF}{Complementary Cumulative Distribution Function}
\newacronym{CDMA}{CDMA}{Code Division Multiple Access}
\newacronym{CTMC}{CTMC}{Continuos-Time Markov Chain}
\newacronym{CP}{CP}{Control Plane}
\newacronym{CQI}{CQI}{Channel Quality Indicator}
\newacronym{CU}{CU}{Centralized Unit}

\newacronym{DL}{DL}{downlink}
\newacronym{DNC}{DNC}{Deterministic Network Calculus}
\newacronym{DRP}{DRP}{Dynamic Resource Provisioning}
\newacronym{DRL}{DRL}{Deep Reinforcement Learnning}
\newacronym{DU}{DU}{Distributed Unit}

\newacronym{eMBB}{eMBB}{enhanced Mobile Broadband}
\newacronym{ETSI}{ETSI}{European Telecommunication Standards Institute}
\newacronym{EBB}{EBB}{Exponentially Bounded Burstiness}
\newacronym{EBF}{EBF}{Exponentially Bounded Fluctuation}
\newacronym{E2E}{E2E}{End-to-End}
\newacronym{EDF}{EDF}{Earliest Deadline First}
\newacronym{EM}{EM}{Expectation-Maximization}

\newacronym{FCFS}{FCFS}{First-come First-served}
\newacronym{FIFO}{FIFO}{First In First Out}

\newacronym{GBR}{GBR}{Guaranteed Bit Rate}
\newacronym{GMM}{GMM}{Gaussian Mixture Model}
\newacronym{GSMA}{GSMA}{Global System for Mobile Communications Association}
\newacronym{GST}{GST}{Generic Network Slice Template}
\newacronym{gNB}{gNB}{Next generation NodeB}

\newacronym{HDR}{HDR}{High Data Rate}

\newacronym{ITU}{ITU}{International Telecommunication Union}
\newacronym{IoT}{IoT}{Internet of Things}

\newacronym{LA}{LA}{Link Adaptation}
\newacronym{LOS}{LOS}{Line-of-sight}
\newacronym{LSTM}{LSTM}{Long Short-Term Memory}
\newacronym{LTE}{LTE}{Long Term Evolution}

\newacronym{MAC}{MAC}{Medium Access Control}
\newacronym{MCS}{MCS}{Modulation and Coding Scheme}
\newacronym{MDN}{MDN}{Mixture Density Network}
\newacronym{MGF}{MGF}{Moment Generating Function}
\newacronym{ML}{ML}{Machine Learning}
\newacronym{MNO}{MNO}{Mobile Network Operator}
\newacronym{mMTC}{mMTC}{Machine Type Communication}
\newacronym{MSE}{MSE}{Mean Squared Error}
\newacronym{mURLLC}{mURLLC}{massive ultra-Reliable Low Latency Communication}

\newacronym{NE}{NE}{Nash Equilibrium}
\newacronym{NEST}{NEST}{Network Slice Type}
\newacronym{NFMF}{NFMF}{Network Function Management Function}
\newacronym{NFV}{NFV}{Network Function Virtualization}
\newacronym{NG-RAN}{NG-RAN}{Next Generation - RAN}
\newacronym{NLOS}{NLOS}{Non-Line-of-sight}
\newacronym{NN}{NN}{Neural Network}
\newacronym{NSO}{NSO}{Network Slice Orchestrator}
\newacronym{NSMF}{NSMF}{Network Slice Management Function}
\newacronym{NSSMF}{NSSMF}{Network Slice Subnet Management Function}
\newacronym{NR}{NR}{New Radio}

\newacronym{OFDMA}{OFDMA}{Orthogonal Frequency-Division Multiple Access}
\newacronym{O-RAN}{O-RAN}{Open RAN}
\newacronym{PDF}{PDF}{Probability Density Function}
\newacronym{PMF}{PMF}{Probability Mass Function}
\newacronym{PRB}{PRB}{Physical Resource Block}
\newacronym{P-NEST}{P-NEST}{private NEST}

\newacronym{QoS}{QoS}{Quality of Service}

\newacronym{RAN}{RAN}{Radio Access Network}
\newacronym{RB}{RB}{Resource Block}
\newacronym{RBG}{RBG}{Resource Block Group}
\newacronym{RIC}{RIC}{RAN Intelligent Controller}
\newacronym{RRM}{RRM}{Radio Resource Management}
\newacronym{RSRP}{RSRP}{Received Signal Received Power}
\newacronym{RSRQ}{RSRQ}{Received Signal Received Quality}
\newacronym{RSSI}{RSSI}{Received Signal Strength Indication}
\newacronym{RT}{RT}{Real Time}
\newacronym{RU}{RU}{Radio Unit}

\newacronym{SDO}{SDO}{Standards Developing Organization}
\newacronym{SINR}{SINR}{Signal-to-Interference-plus-Noise Ratio}
\newacronym{SLA}{SLA}{Service Level Agreement}
\newacronym{SNC}{SNC}{Stochastic Network Calculus}
\newacronym{S-NEST}{S-NEST}{standardized NEST}

\newacronym{TTI}{TTI}{Transmission Time Interval}

\newacronym{UE}{UE}{User Equipment}
\newacronym{UL}{UL}{Uplink}
\newacronym{UP}{UP}{User Plane}
\newacronym{uRLLC}{uRLLC}{ultra-Reliable Low Latency Communication}

\newacronym{V2X}{V2X}{Vehicle-to-Everything}
\newacronym{VBR}{VBR}{Variable Bit Rate}
\newacronym{VR}{VR}{Virtual Reality}
\newacronym{vRAN}{vRAN}{virtualized RAN}
\newacronym{vBS}{vBS}{virtualized Base Station}

\newacronym{WiMAX}{WiMAX}{Worldwide Interoperability for Microwave Access}
\newacronym{WCDMA}{WCDMA}{Wideband \gls{CDMA}}